	\documentclass[journal, onecolumn]{IEEEtran}

	\usepackage{graphicx}          % Include this line if your document contains figures,
	\usepackage[font=scriptsize,labelfont=bf]{caption}
	\usepackage{tikz} 
	\usetikzlibrary{shapes.geometric}
	\usetikzlibrary{arrows.meta, arrows, automata, positioning, shapes, backgrounds}
	\usetikzlibrary{calc}
	\usetikzlibrary{circuits.ee.IEC}
	\usetikzlibrary{chains}

	\usepackage{textcomp}
	\usepackage{textcomp}

	\usepackage{algorithm}
	\usepackage[noend]{algpseudocode}

	\usepackage{amssymb}
	\usepackage{amsmath}
	\usepackage{bigints}
	\usepackage{enumerate}% http://ctan.org/pkg/enumerate
	\usepackage{xcolor}
	
	\usepackage{pgfplots}
	\usepackage{caption}
	\usepackage{subcaption}
	\usepackage{dsfont}
	\usepackage[colorlinks=false,linkcolor=gray]{hyperref}
	
	\makeatletter
	\renewcommand*{\eqref}[1]{%
		\hyperref[{#1}]{\textup{\tagform@{\ref*{#1}}}}%
	}
	\makeatother
	
	\newtheorem{remark}{\textbf{Remark}} %[in_counter]
	\newtheorem{assumption}{\textbf{Assumption}}
	
	\newcommand\numberthis{\addtocounter{equation}{1}\tag{\theequation}}
	\newtheorem{lemma}{\textbf{Lemma}} %[in_counter]
	\newtheorem{theorem}{\textbf{Theorem}} %[in_counter]
	\newtheorem{proposition}{\textbf{Proposition}} %[in_counter]
	\newtheorem{problem}{\textbf{Problem}} %[in_counter]
	\DeclareMathAlphabet{\mathpzc}{OT1}{pzc}{m}{it}
\def\@fnsymbol#1{\ensuremath{\ifcase#1\or *\or \dagger\or \ddagger\or
		\mathsection\or \mathparagraph\or \|\or **\or \dagger\dagger
		\or \ddagger\ddagger \else\@ctrerr\fi}}

\usepackage[many]{tcolorbox}

	\newtcolorbox{cross}{blank,breakable,parbox=false,
		overlay={\draw[red,line width=5pt] (interior.south west)--(interior.north east);
			\draw[red,line width=5pt] (interior.north west)--(interior.south east);}}
	
\begin{document}
%	\title{Distributed Value of Information in Multi-hop Networks}
	%\title{Distributed Value of Information in Multi-loop Multi-hop Networked Control}
	%\title{Distributed Value of Information in Multi-hop Networked Multi-loop Control}
	\title{Distributed Value of Information in Feedback Control over Multi-hop Networks}
	
	\author{Precious~Ugo~Abara and Sandra~Hirche,~\IEEEmembership{Fellow,~IEEE}% <-this % stops a space
	\thanks{Precious Ugo~Abara and Sandra~Hirche are with Technical University of Munich, Germany,
		Department of Electrical and Computer Engineering,
		Chair of Information-oriented Control (ITR) e-mail: \{ugoabara, hirche\}@tum.de.}
	\thanks{This work has been submitted to the IEEE for possible publication. Copyright may be transferred without notice, after which this version may no longer be accessible.}
	\thanks{This work has been funded by the German Research Foundation (DFG) under the grant number 315177489 as part of the SPP 1914 (CPN).}% <-this % stops a space
	}

	\markboth{Preprint}{Ugo~Abara \MakeLowercase{\textit{et al.}}}

	\maketitle
	
	\begin{abstract}
		Recent works in the domain of networked control systems have demonstrated that the joint design of medium access control strategies and control strategies for the closed-loop system is beneficial. However, several metrics introduced so far fail in either appropriately representing the network requirements or in capturing how valuable the data is. In this paper we propose a distributed value of information (dVoI) metric for the joint design of control and schedulers for medium access in a multi-loop system and multi-hop network. We start by providing conditions under certainty equivalent controller is optimal. Then we reformulate the joint control and communication problem as a Bellman-like equation. The corresponding dynamic programming problem is solved in a distributed fashion by the proposed VoI-based scheduling policies for the multi-loop multi-hop networked control system, which outperforms the well-known time-triggered periodic sampling policies. Additionally we show that the dVoI-based scheduling policies are independent of each other, both loop-wise and hop-wise. At last, we illustrate the results with a numerical example.
	\end{abstract}
	
	\begin{IEEEkeywords}
	Distributed Value of Information, Multi-loop Multi-hop Dynamic Programming, Scheduling Problem Decomposition
	\end{IEEEkeywords}

	% \fi
	\IEEEpeerreviewmaketitle

	\section{Introduction}
	\IEEEPARstart{I}{}n recent years technology development has been moving towards networked control systems. In such domain, a plethora of different agents and their respective feedback loops are constrained to communicate through a shared network. The application domains are numerous and include for example robotics \cite{fan2020event}, smart energy grids \cite{wang2021event}, autonomous driving \cite{Ploger2018}, and smart cities \cite{doi:10.1177/0020720920983503}. From a theoretical perspective the common denominator of the aforementioned applications is the presence of multiple feedback loops closed over a shared communication network. 
	{In such a context, the two main layers of the system – control and communication – have a significant impact on each other's performance and must deal with heterogeneous and time-varying limitations and demands. As a result, the overall system design necessitates novel cross-layer joint design techniques that are adaptable and scalable, as well as responsive to real-time alterations in individual layers.}
	
	{At its most basic level, a controlled cyber-physical system is a feedback system in which sensory data is transmitted to a controller via a communication channel in order to adjust the system's behavior in a desirable manner. As a result, the controller's ability to control the system is limited by the information available it has access to.} In classical control theory, observations of a dynamical process are sampled equidistantly in time and transmitted directly to the controller. In virtue of the seminal work \cite{astrom2002}, it has become unambiguous that not all observations of a given process have the same importance. {This validates the control community's growing interest in event-triggered control, i.e., control schemes where the observations are transmitted only when deemed to be important according to some metric.} Event-triggered control takes a central role in networked control systems. In fact, in such systems some challenges and constraints - such as control objectives minimization, energy constraints, limited bandwidth, and contention resolution etc. - arise naturally. The agents must therefore reach an agreement so that access to the communication medium can be regulated while simultaneously ensuring their control objectives are met. {The necessity for medium access distributed coordination, on the other hand, stems from the intrinsic allocation difficulty posed by the agents' heterogeneous and dynamic character.}
	%The allocation problem of limited communication resources is a challenging research area in the design of networked control systems. 
	%In a conflict-free protocol the channel is assigned to the users without any overlap between them. Time division multiple access (TDMA), and code division multiple access (CDMA) are two common centralized protocols often preferred in medium-size networks. Such protocols are therefore collision-free and offer fairness among agents, due to the precise channel scheduling. However, static conflict-free protocols do not utilize the shared channel very efficiently  due to the allocation being predetermined at the network design time. Precisely, when the loads of the agents are asymmetric the channel might be idle even though some agents have data to transmit. 
	In a dynamic allocation strategy, the channel allocation changes with time and is based on current and event-dependent demands of the various agents. {A dynamic strategy achieves greater responsiveness and better channel resource utilization at the expense of additional control overhead, which is unnecessary with static allocation strategies and typically consumes a portion of the channel \cite{capone2006}. However, when communication constraints and stochastic control processes are taken into account, the benefits of using dynamic allocation strategies surpass those of static allocation strategies \cite{brunner2018}}. 
	%Deterministic scheduling policies usually render improved performance in comparison to randomized ones as they award the channel to systems with the highest priority \cite{wu2013}. However, they often lack scalability and flexibility in dealing with channel imperfections. 
	
	In the common setting of a single-hop communication network, event-triggered control systems have been shown to inherently possess medium access arbitration mechanism for networked control systems \cite{astrom2002, Astrom2008}, while guaranteeing at the same time local control performances \cite{molin2013}. As a consequence, a significant reduction of data transmission can be obtained without degrading the overall control performance. A non-exhaustive list of interesting works exploiting event-triggering mechanism in single-hop networks include \cite{MAMDUHI2017209, Dimarogonas2012, Liu2021}, and \cite{Zhang2020}. As discussed previously, the joint design of communication and control systems provides flexible allocation strategies and improves control performances in the presence of constraints. However, finding joint optimal strategies is in general a hard problem even for a single-hop network \cite{witsenhausen1986}. This is due to the fact that the event-trigger and the controller may have non-classical information pattern, and the arrival of data packets from the event-trigger may be influenced by the controller \cite{ramesh2011}. {This complicates the derivation of optimal policies significantly since linear control approaches may not be optimal. To make matters worse, for real-world applications, the design technique must be scalable and adaptable to varied system configurations.}
	
	In this work we are concerned with the joint design of optimal scheduling strategies and control policies in a multi-loop networked control system over a shared multi-hop communication channel. {Multi-hop transmission networks have been considered as an effective way to increase the coverage, throughput and transmission reliability \cite{sahoo2017}.} 
	In a multi-loop system with $N$ sub-systems and multi-loop network with $L$ hops, the increased difficulty can be seen by the fact that optimal policies for $N(L+1)$ decision makers must be designed. 
	On top of it all, in a multi-hop network the complete history of the control actions is not available at the schedulers, as opposed to a single-hop network.
	Furthermore, the tractability of the problem depends on the information pattern of all decision makers. {In this article, we introduce a novel distributed value-of-information-based scheduling technique for the coupled communication and control performance maximization problem that is adaptable to varied system configurations and scalable in a multi-loop multi-hop system.} To the best of our knowledge this is the first time such problem setting has been considered in literature. %However, the design of event-triggers and controllers that yield the optimal control-communication trade-off for heterogeneous systems in a multi-hop network is still an open problem.
	
	%Observing that the problem in \eqref{eq: dual problem} is strongly related the network utility maximization problem \cite{molin2014, shakkottai2008network}, in what follows we will derive a \textit{utility-of-information-based} scheduling policy that solves \eqref{eq: dual problem}
	\subsection{Related Work}
	{
	When merging feedback control and medium access, there are numerous challenges to overcome. Despite the fact that in a single-loop single-hop configuration, the decision makers participating in the optimization problem are only one scheduler and one controller, the challenge of joint communication and control design remains intractable in general (see e.g., \cite{witsenhausen1986}\cite{wu2013}). The rationale is that the optimal estimator at the controller might be nonlinear and have no analytical solution, estimation and control are coupled because of the dual effect, and schedulers and controllers have non-classical information patterns. The authors in \cite{ramesh2011} studied dual effect in event-triggered control, and proved that it exists in general. In addition, they show for a single-loop single-hop system that the certainty equivalence principle holds if and only if the triggering policy is independent of the control policy, i.e., the arrival of a data packet is uncorrelated to the plant state.
	In a similar way, \cite{demirel2019} propose an optimal event-triggered control with imperfect information that used a stochastic triggering policy independent of the control policy and proved that the optimal control policy is a certainty-equivalence policy. 
	
	 In their study of single-loop single-hop systems, the authors in  \cite{RAMESH2009328} analyze the impact of a medium access controller on quadratic cost functions, and found that the decision statistics of the scheduling strategy are responsible for an additive increase in cost. Furthermore, they design a threshold-based adaptive scheduling strategy, and verify through simulations, that it reduces the increase in cost associated with the decision strategy. However, as the authors conclude, finding the optimal adaptive method remains a challenge, especially when the problem is extended to include multiple loops over a shared network.
	%In \cite{araujo2014} a dynamic utilization policy is identified for the TDMA slots of the IEEE802.15.4 protocol which minimizes the network bandwidth usage for a single-loop.
	
	%Moving towards a more systematic approach, 
	In recent works \cite{kosta2017age, tourajage2018, Kompella2019} the metric age of information (AoI)  characterizing information staleness is studied. The AoI metric is defined as the temporal gap between current time and the generation time of the latest received signal. The authors in \cite{tourajage2018} analyzed the effect of information staleness in a single-loop single-hop. They showed that there is a trade-off between control performance and information staleness, i.e., medium access control, in the joint optimization problem. 
	%In \cite{Bedewy2019} the authors proposed a policy which minimizes the age of information in a multi-hop network. 
	However, as it is shown in \cite{ayan2019}, AoI is not the only determining parameter in networked control systems  - in particular when there are heterogeneous tasks requirements and stochastic system dynamics. %Specifically, in control problems the AoI metric alone is not able to characterize different task criticalities.
	
	Preliminary works in the direction of joint design of scheduling and control policies were presented in \cite{soleymani2018value, MOLIN2019108578}. The authors in \cite{soleymani2018value} introduced the notion of value of information (VoI). They quantified the VoI as the variation in the cost-to-go of the system given an observation at a certain time instant. Furthermore, they developed a theoretical framework for the joint design of an event trigger and a controller in optimal event-triggered control for a single-loop system. Moreover, they characterized the VoI in the trade-off between the sampling rate and control performance, and synthesized a closed-form suboptimal VoI-based scheduling policy with a performance guarantee. In \cite{MOLIN2019108578}, the authors developed a heuristic-based data scheduling methodology for a system of networked estimators which uses the VoI to determine when data must be transmitted over a shared communication network. The proposed prioritizing policy was theoretically shown to improve remote estimation performance for decoupled estimators.	However, both works \cite{soleymani2018value, MOLIN2019108578} assumed a single-hop communication channel; the extension to multi-hop networks is not straightforward.}	
	
	In multi-loop single-hop networked control systems, attempts have been focused on designing scheduling mechanisms such that the current status of the control systems are taken into account when arbitrating the channel access \cite{Mamduhi2014EventbasedSO, molinramesh2015}. In \cite{molinramesh2015}, for example, an heuristic  policy for an innovations-based priority scheme for scheduling multiple sensor data over a CAN-like networked control system is proposed. A  distinct feature of the approach is that each sensor computes the value of information based on its own available information. The future scheduling decisions are predetermined by a baseline heuristic, in order to simplify the computation of the value of information. In \cite{Mamduhi2014EventbasedSO}, the authors proposed an heuristic prioritized error-based measure. In this approach the scheduler allocates the communication resources based on the prioritized error-based measure and the control policy is assumed to be known a priori. To the best of our knowledge, all previous works in the direction of joint design of schedulers and controller consider either a single-loop single-hop networked system or multi-loop single-hop networked system. Against this background, we postulate that finding the optimal medium access and control performance in a multi-loop multi-hop networked control system is still an open problem.

	\subsection{Contributions}
	Targeting the shortcomings of existing design approaches for medium access and control performance coordination in networked control systems, in this article we propose a distributed value of information metric that aims at serving different task criticalities in a multi-loop multi-hop networked control system. The asynchronous policy makers adaptively accommodate for dynamically evolving network and control requirements. Additionally we show that the proposed dVoI inherits the scalability feature of the AoI metric while sufficiently accounting for individual task requirements of closed control loops. Henceforth, the main contribution are summarized as follows:
	\begin{enumerate}[i.]
		\item We show that under a specific class of scheduling policies the certainty equivalence principle holds in a multi-loop multi-hop networked system.
		\item We reformulate the joint scheduling and control problem as an equivalent Bellman-like equation, i.e., a two-dimensional Bellman equation. 
		\item Using rollout policy approximation we decompose the Bellman-like equation into independent sub-problems of the same form. Additionally, we show that the approximation outperforms periodic time-triggered policy.
		\item We define the distributed value of information between any two neighboring decision makers $j$  and $j+1$ as the measure of the quality of data between the them. Using this metric we synthesized a closed-form dVoI-based scheduling with guaranteed performance. In other words, whenever the dVoI between them is negative then the quality of data is not satisfactory and must be restored by sending updated information from $j$ to $j+1$.
		\item Motivated by the fact that the complete history of the control actions is not available at the schedulers, we proposed and discussed a solution to the problem of joint scheduling and control in the presence of unknown inputs at the schedulers. 
	\end{enumerate}

	%The remainder of this article is organized as follows. In Section~\ref{sec: problem statement} we introduce the system model and describe the problem setting.
	\subsubsection*{Notation}
	In this article, the operator $\left(\cdot\right)^\top$ denotes the transpose. The expectation operator is denoted by ${\mathbb{E}}\left[\cdot\right]$, and with ${\mathbb{E}}\left[\cdot\Big\vert \cdot\right]$ the conditional expectation, and ${\rm tr}(\cdot)$ denotes the trace operator. The Euclidean norm is denoted by $\Vert \cdot \Vert_2$. 
	
	%\IEEEpubidadjcol
	
	\section{Preliminaries}
	\label{sec: problem statement}
	\subsection{System Model}
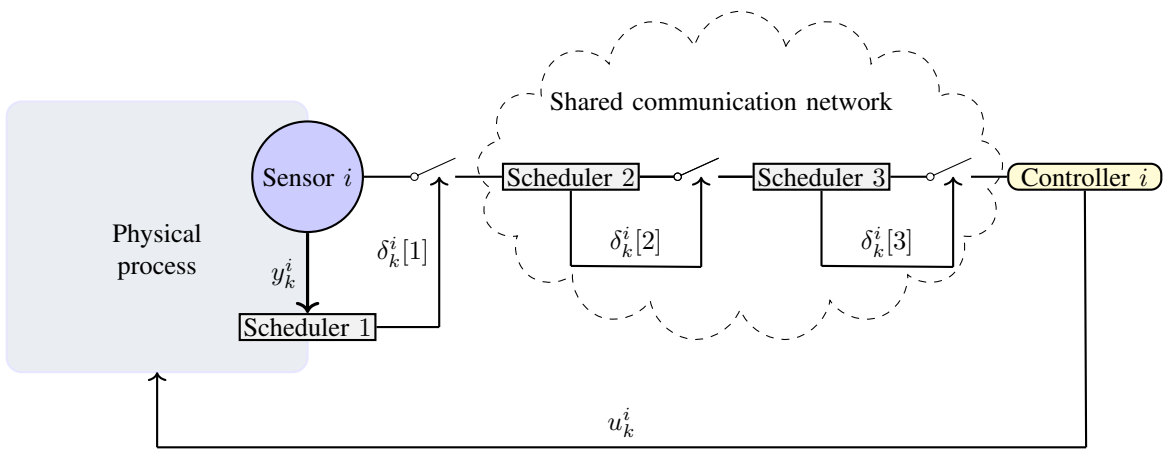
\begin{figure}[h!]
	\centering
\begin{tikzpicture}[start chain=going right,
estimator/.style={draw,thick,rounded corners,fill=yellow!20,inner sep=.05cm},
process/.style={draw,thick,circle,fill=blue!20, inner sep=0.1},circuit ee IEC,
state/.style={draw,thick,fill=gray!10,inner sep=.05cm}]
\tikzset{%
	in place/.style={
		auto=false,
		fill=white,
		inner sep=2pt,
	},
}
\draw[blue, thick,fill={rgb: red,0.31; green, 0.63; blue, 0.87}, opacity=0.1, rounded corners=.2cm] (-4,1) rectangle (0,-2.6);
\coordinate (GP) at (-2, -1);

\node[align=center] at (GP) {Physical\\process};

\node[process,on chain] (0) {\ Sensor $i$\ \ };
\node[state,on chain, left of=0, xshift=3.5cm] (1)  {Scheduler $2$};
%\node[state,on chain] (2)  {Scheduler $3$};
\node[state,on chain, xshift=0.5cm] (2) {Scheduler $3$};
\node[estimator,on chain, left of=2, xshift=3.5cm] (3) {\ Controller $i$\ \ };
\node[state, below of=0, yshift=-1cm] (4) {Scheduler $1$};
\draw[->,very thick] (0) -- node[left] {$y_k^i$} (4);

\coordinate[below=1.6cm of GP] (process-ap);
\coordinate[below=1cm of process-ap] (process-ap-1);
\coordinate[right=12.35cm of process-ap-1] (controller-1);
\draw[-,thick] (process-ap-1) -- node[above] {$u_k^i$} (controller-1);
\draw[-,thick] (3) --  (controller-1);
\draw[->,thick] (process-ap-1) -- (process-ap);

% event-trigger switch
\coordinate (below-first-switch) at  ($(0)!0.5!(1)$);
\coordinate[below=2cm of below-first-switch] (out-event-trigger);

\draw[->,thick] (out-event-trigger) -- node[left] {$\delta_{k}^i[1]$} (below-first-switch);
\draw[-,thick] (4) --  (out-event-trigger);

\draw[set make contact graphic = var make contact IEC graphic, thick] (0) to [make contact={info ={$ $}, info'={$ $}}] (1) ;

% hop 1 switch		
\draw[set make contact graphic = var make contact IEC graphic, thick] (1) to [make contact={info ={$ $}, info'={$ $}}] (2) ;

\coordinate (switch-hop1) at  ($(1)!0.52!(2)$);
\coordinate[below=1cm of 1] (out-hop1);
\coordinate[below=1.2cm of switch-hop1] (out-switch-hop1);
\draw[-,thick] (1) --  (out-hop1);
\draw[-,thick] (out-hop1) -- node[above] {$\delta_{k}^i[2]$} (out-switch-hop1);
\draw[->,thick] (out-switch-hop1) -- (switch-hop1);

% hop 2 switch		
\draw[set make contact graphic = var make contact IEC graphic, thick] (1) to [make contact={info ={$ $}, info'={$ $}}] (2) ;

% hop 3 switch		
\draw[set make contact graphic = var make contact IEC graphic, thick] (2) to [make contact={info ={$ $}, info'={$ $}}] (3) ;

\coordinate (switch-hop2) at  ($(2)!0.5!(3)$);
\coordinate[below=1cm of 2] (out-hop2);
\coordinate[below=1.2cm of switch-hop2] (out-switch-hop2);
\draw[-,thick] (2) --  (out-hop2);
\draw[-,thick] (out-hop2) -- node[above] {$\delta_{k}^i[3]$} (out-switch-hop2);
\draw[->,thick] (out-switch-hop2) -- (switch-hop2);	

\node[cloud, cloud puffs=17, cloud ignores aspect, minimum width=7cm, minimum height=4.4cm, align=center, draw, dashed] (Pcloud) at (5.75cm, 0cm) {};
\node[align=center] at (5.5,1) {Shared communication network};
\end{tikzpicture}
\caption{
	Example of a $3$-hop network. The first scheduler for each system $i = 1, \ldots, N$ decides whether to transmit the measurement $\hat{x}_k^i$ at the first hop by appropriately choosing $\delta_{k}^i[1]$. The two network schedulers are responsible of deciding $\delta_{k}^i[2], \delta_{k}^i[3]$ for system $i$. Moreover, the communication network is shared among all $N$ loop. } 
\label{fig: physical setup}
\end{figure}

	Consider $N$ discrete-time linear process
	\begin{align}
	\label{eq: global system}
	\begin{cases}
	{x}_{k+1}^i & = {A_i}{x}_k^i + B_i u_k^i + w_k^i \\
	\quad {y}_k^i & = {C_i} {x}_k^i + v_k^i\\
	\end{cases}
	\end{align}
	where ${x_k^i} \in \mathbb{R}^{n_i}$ is the state of the system, $A_i \in \mathbb{R}^{n_i \times n_i}$, $B_i \in \mathbb{R}^{n_i \times p_i}$, and $C_i \in \mathbb{R}^{m_i \times n_i}$. The process noise ${w}_k^i \in \mathbb{R}^{n_i}$ and measurement noise ${v}_k^i \in \mathbb{R}^{m_i}$ are characterized as unbiased independent Gaussian random vectors, i.e, $w_k^i \sim \mathcal{N}(0, W_i)$ and $v_k^i \sim \mathcal{N}(0, V_i)$ for some positive semidefinite $W_i$ and positive definite $V_i$.
	The initial state $x_0^i$ is a random variable, with zero mean and finite covariance $\Omega_0^i$, which is assumed to be statistically independent of the process noise $w_k^i$ and measurement noise $v_k^i$ for all $k$ and for all $i$. 	
	In addition to the physical processes, sensors and controllers, the system also consists of a shared multi-hop communication network (see Fig.~\ref{fig: physical setup}).
%	\begin{figure}[h!]
%			\centering
%			{\includegraphics[trim=2.5cm 8cm 8.5cm 4.2cm,width=0.48\textwidth]{images/networknew.pdf}}
%			\caption{The measurement $y_k^i$ is sent to the estimator according to the event-triggering criteria. Moreover, the communication network is shared among all $N$ loop. The dashed line from the estimator to the event-trigger indicates that we assume the event-trigger can compute $\hat{y}_{k \vert k -1} ^i$ without actual communication.} 
%			\label{fig: physical setup}
%	\end{figure}
	\subsection{Network Model} 	
	\subsubsection*{Multi-hop}
	We assume the communication network to be a multi-hop network with $L>0$ hops, where each hop $j = 1, \ldots, L$ is equipped with a tuple of schedulers $\mathcal{S}[j] = \left(\mathcal{S}^1[j], \ldots, \mathcal{S}^N[j]\right)$ responsible of the choices $\left(\delta_{k}^1[j], \ldots, \delta_{k}^N[j] \right)$ for all $k$. The scheduling variable $\delta_{k}^i[j]$ indicates whether the information from system $i$ should be forwarded at hop $j$ and time $k$ (see Fig.~\ref{fig: physical setup}). For each system $i\in \{1, \ldots, N\}$, the first scheduler, i.e., $\mathcal{S}^i[j]$, is co-located with the sensor of the system. It is then clear that our problem setting can be seen as a set of $N (L+1)$ decision makers, i.e., $N*L$ schedulers and $N$ controllers. Furthermore, for a system $i\in \{1, \ldots, N\}$, we indicate with $\mathcal{H}[i]$ the chain formed by the decision makers from the $i$-th sensor to the $i$-th controller
	\begin{align}
		\label{eq: chain DMs}
		\mathcal{H}[i] = \left(\mathcal{S}^i[1], \ldots, \mathcal{S}^i[L], \mathcal{C}^i\right),
	\end{align} 
	where $\mathcal{C}^i$ is the controller of the $i$-th system.
	\begin{remark}
		
		\label{rem: different number of hops}
		For notational convenience, we assumed that each control loop is closed through a multi-hop network with a fixed number of hops $L$. As will be discussed later, this assumption is not restrictive since our approach and results are readily extensible to the case where the control loops $i=1,\ldots, N$ are closed through a different number of communication hops $L_i$.
	\end{remark}
	\subsubsection*{Communication Delay at Decision Makers}
	For any system $i = 1, \ldots, N$, consider the $(L+1)$-tuple of decision makers $\mathcal{H}[i]$ defined in \eqref{eq: chain DMs}. For $j = 1, \ldots, L+1$ we assume a constant delay $d[j]$ between decision maker $j$ and $j-1$, which is also independent of the system $i$. In other words, the data from decision maker $j-1$ is available with at least $d[j]$ steps delay to decision maker $j$ for any system. Additionally, we assume $d[1] = 0$, i.e., measurements from a sensor are immediately available to the first scheduler in the decision chain. The total delay at a decision maker $j \in \{1, \ldots, L+1\}$ is therefore
	\begin{align}
		\label{eq: total delay at j}
		{D}[j] = \sum_{l=1}^{j} d[l].
	\end{align}
	\section{Information Structure}
	\begin{assumption}
		\label{ass: broadcast of estimates instead of states}
		For each system $i\in \{1, \ldots, N\}$ and at each time instant $k$, scheduler $\mathcal{S}^i[1]$ receives $y_k^i$ with zero-steps delay, computes the state estimate $\hat{x}_{k\vert k}^i[1]$, and decides whether to transmit $\hat{x}_{k\vert k}^i[1]$ to the next decision maker in the chain $\mathcal{H}[i]$ by appropriately choosing the scheduling variable $\delta_{k}^i[1]$.
	\end{assumption}
	We now give two groups of information sets.
	\subsubsection*{Schedulers}
	For each system $i = 1,\ldots, N$, the information available to scheduler $\mathcal{S}^i[j]$ at hop $j>1$ set is
	\begin{subequations}
		\label{eq: information set schedulers}
		\begin{equation}
		\label{eq: information set hop l}
		\mathcal{J}_{k}^i[j] = \left\lbrace \mathcal{J}_t^i[j-1] \Big\vert \delta_{t-d[j]}^i[j-1] = 1, t \leq k\right\rbrace \cup \left\lbrace\delta_{k-d[j]}^i[j-1], \delta_{0:k-1}^i[j] \right\rbrace, \quad \text{for }j = 2, \ldots, L.
		\end{equation} 
	Given that the state estimate $\hat{x}_{k\vert k}^i[1]$ is a sufficient statistic for the conditional distribution of $x_k^i$ given the measurement history \cite{Astrom2008}, from Assumption~\ref{ass: broadcast of estimates instead of states} the information set $\mathcal{J}_{k}^i[1]$ at scheduler $\mathcal{S}^i[1]$ can be written as
		\begin{equation}
		\label{eq: information set event-trigger}
		\mathcal{J}_{k}^i[1] = \left\lbrace \hat{x}_{t\vert t}^i[1] \Big\vert t \leq k \right\rbrace \cup \left\lbrace \delta_{0:k-1}^i[1] \right\rbrace.
		\end{equation} 
	The total information available at hop $j=1, \ldots, L$ is $\{\mathcal{J}_{k}^1[j], \ldots, \mathcal{J}_{k}^N[j]\}$.
	%We observe that \eqref{eq: information set event-trigger} and \eqref{eq: information set hop l} requires the broadcast of $u_k^i$ both to the event-triggers and the network schedulers. However, we will see that this overhead can be avoided.
	\subsubsection*{Controllers}
	Due to the presence of schedulers in the multi-hop communication channel, the available information set $\mathcal{I}_k^i$ at the $i$-th controller will be composed only of successfully received state estimates, i.e. not dropped by any of the schedulers. Formally, at sampling time $k$ the information set at the $i$-th controller, for $i=1,\ldots,N$ is
	\begin{equation}
	\label{eq: information set estimator}
	\mathcal{I}_k^i = \left\lbrace \mathcal{J}_t^i[L] \Big\vert \delta_{t-d[L+1]}^i[L] = 1, t \leq k\right\rbrace \cup \left\lbrace\delta_{k-d[L+1]}^i[L], u_{0:k-1}^i \right\rbrace.
	\end{equation}
	\end{subequations}
	In addition to the information sets given in \eqref{eq: information set schedulers}, the controllers and schedulers know the system matrices $A_i, B_i, C_i$ and the noise covariance matrices $V_i, W_i$ of their respective control loop.
	\section{Problem Statement}
	{
	In this article we investigate the joint design of optimal scheduling strategies and control policies in a multi-loop networked control system over a shared multi-hop communication channel. Generally speaking, such a joint global optimization problem is complex to solve in a distributed fashion due to possibly non-classical information pattern \cite{witsenhausen1986}, and large number of control loops and network hops. Stochastic control problems with non-classical information pattern generally do not allow to apply concepts like dynamic programming directly. In this section we introduce the novel distributed value of information (dVoI) metric for the joint design of control and scheduling policies in multi-loop multi-hop networked control systems. As the global performance metric is the aggregation of local control objective functions and communication constraints, the ultimate goal is to guarantee a certain level of performance of all control loops under the designed dVoI policies. %In particular, we are interested in the finding optimal policies for the controllers and the schedulers which guarantees certain global performance metrics. Under some assumptions we provide distributed optimal event-triggered scheduling strategies and control policies in a multi-loop multi-hop networked control system.
} %The existence of stationary policies is guaranteed under some mild conditions \cite{bertsekas1995dynamic, puterman2014markov}.
	%A stationary control policy is an admissible policy of the form $(\mu^i, \mu^i, \ldots)$. We refer to $(\mu^i, \mu^i, \ldots)$ as $\mu^i$, and define the augmented vector of control policies $\mu \triangleq (\mu^1, \ldots, \mu^N)$.
	%A stationary scheduling policy is an admissible policy of the form $(\pi^i[j], \pi^i[j], \ldots)$. We refer to $(\pi^i[j], \pi^i[j], \ldots)$ as $\pi^i[j]$, and define the augmented vector of scheduling policy policies $\pi [j] \triangleq (\pi^1[j], \ldots, \pi^N[j])$, and $\pi = (\pi[0], \ldots, \pi[L])$.
	\subsection{Control Performance}
	Regarding the control objectives, we choose the standard LQG finite cost function
	\begin{align*}
	\numberthis \label{eq: control cost}
	J^i(\mu, \pi) = \mathbb{E}\left[(x_T^i)^\top {\Lambda_{i}} x_T + \sum_{k=0}^{T-1} (x_k^i)^\top Q_i x_k + (u_k^i)^\top R_i u_k^i \right],
	\end{align*}
	under the constraint $u_k^i = \mu_k^i(\mathcal{I}_{k}^i)$. The matrix $Q_i$ is semi-definite positive, meanwhile $R_k^i$ is definite positive, the pair $(A_i, B_i)$ is stabilizable and $(A_i, Q_i^{\frac{1}{2}})$ is detectable, with $Q_i = \left(Q_i^{\frac{1}{2}}\right)^\top Q_i^{\frac{1}{2}}$. The control policy $\mu_k^i$ is a causal, measurable and admissible function of the available information at the controller. Formally, for i = 1, $\ldots, N$
	\begin{align}
	\label{eq: control policy}
	u_k^i = \mu_k^i\left(\mathcal{I}_k^i\right),
	\end{align}
	where the information set $\mathcal{I}_k^i$ is defined in \eqref{eq: information set estimator}. We refer to $(\mu^i_0, \ldots, \mu^i_{T-1})$ as $\mu^i$, and define the augmented vector of control policies $\mu \triangleq (\mu^1, \ldots, \mu^N)$.
	\subsection{Communication Constraints}
	Since several systems share the same network at time $k$, the network manager imposes a communication constraint at each network scheduler $j \in \{1, \ldots, L\}$. As it will take a central role in what follows we define the individual request rate $r^i[j]$ of the $i$-th system at the network scheduler $j$ by
	\begin{subequations}
		\begin{align*}
		\numberthis \label{eq: individual request rate}
		r^i[j] = \mathbb{E} \left[\sum_{k=0}^{T - 1}  \delta_k^i[j]\right].
		\end{align*}
		Moreover, we define the total request rate at a network scheduler $j$ as
		\begin{align*}
		\numberthis \label{eq: total average request rate}
		r[j] = \sum_{i=1}^{N}r^i[j].
		\end{align*}
	\end{subequations}	
	The constraint imposed by each network scheduler can then be written as
	\begin{align*}
	\numberthis \label{eq: triggering cost}
	r[j] \leq R[j], \qquad j = 1, \ldots, L.
	\end{align*}
	Without loss of generality we assume $R[L] \leq R[L-1] \leq \cdots \leq R[1]$. 
	The scheduling policy $\pi_k^i[j]$ is a causal, measurable and admissible function of the available information at network scheduler $j$. Formally,
	\begin{align}
	\label{eq: triggering policy j}
	\delta_k^i[j] = \pi_k^i[j]\left(\mathcal{J}_k^i[j]\right), \quad j = 1, \ldots, L.
	\end{align}
	We refer to $(\pi^i_0[j], \ldots,  \pi^i_{T-1}[j])$ as $\pi^i[j]$, and define the augmented vector of scheduling policies at hop $j$ as $\pi [j] \triangleq (\pi^1[j], \ldots, \pi^N[j])$, and $\pi \triangleq (\pi[1], \ldots, \pi[L])$.
	\subsection{Global Optimization Problem}
	The joint global optimization problem which we focus on in this article is given by
	\begin{problem}[Global Optimization Problem]
		\label{prob: optimization problem}
		\begin{align*}
		%\numberthis\label{eq: minimization global}
		\underset{\pi, \mu }{\operatorname{min}} \quad & \sum_{i = 1}^{N} J^i(\mu, \pi) \\
		%\label{eq: minimization global constraint}
		\operatorname{subject\ to } \quad & \eqref{eq: triggering cost}.
		\end{align*}
	\end{problem}
{
		 A hard constraint, i.e., of the form  $\sum_{i=1}^{N} \sum_{k=0}^{T - 1}  \delta_k^i[j] \leq R[j]$, for $j=1, \ldots, L$, poses a condition which must be fulfilled by every sample path of the primitive random variables. Instantaneous violations of the condition may lead to contention between the control loops and, consequently, to a high complexity of the policy design \cite{hadi2021}. On the contrary, a soft constraint, as in \eqref{eq: triggering cost}, is a constraint on the expected number of transmission during the time interval $T$ for a given scheduler. In this case, the optimal policy design in the presence of instantaneous violation do not carry increased complexity as long as the overall expected value is within permissible bounds \cite{molin2013}.
}
	\section{Network Estimators}
	\label{sec: network estimators}
	In this section we will characterize the optimal estimators at each decision maker in the decision chain  $\mathcal{H}[i]$ defined in \eqref{eq: chain DMs} for a given $i \in \{1, \ldots, N\}$. For easier analysis we initially assume that history of the control actions $u_{0:k-1}^i$ is available to every scheduler. After the derivation of our dVoI-based scheduling, we will discuss how this assumption can be relaxed.
	\subsection{First Optimal Estimator}
	\label{sec: kalman estimator}
	For a system $i \in \{1, \ldots, N\}$, the optimal estimator at the first decision maker in $\mathcal{H}[i]$ is the standard Kalman filter given in the following.
	\begin{theorem}[Optimal Estimator \cite{kalman1960new}]
		\label{thm: kalman filter}
		For $i = 1, \ldots, N$ let $\hat{x}_{k\vert k}^i[1]$ denote the estimate of $x_k^i$ conditioned on the measurement history $\{y_{0:k}^i, u_{0:k-1}^i\}$, or analogously $\mathcal{J}_{k}^i[1]$. The mean squared error
		is minimized by the Kalman filter
		\begin{subequations}
			\label{eq: kalman filter equations}
			\begin{align*}
			\numberthis\label{eq: kalman filter equations at event-trigger}
			&\hat{x}_{k\vert k - 1}^i[1] = A_i\hat{x}_{k - 1\vert k - 1}^i[1] + B_i u_{k - 1}^i \\
			&\hat{x}_{k\vert k}^i[1] = \hat{x}_{k \vert k-1}^i[1] + \zeta_k^i,
			\end{align*}
			where the innovation $\zeta_k^i$ is 
			\begin{align}
			\label{eq: innovation at zero}
			\zeta_k^i={K}_{k}^i\left(y_{k}^i - C  \hat{x}_{k\vert k-1}^i[1]\right).
			\end{align}
			The optimal Kalman gain is 
			\begin{align*}
			%\numberthis\label{eq: kalman gain}
			&{K}_k^i = \Sigma_{k \vert k-1}^i C^\top_i \left(C_i \Sigma_{k \vert k-1}^i C^\top_i + V_i\right)^{-1},
			\end{align*}
			and error covariance is 
			\begin{align*}
			&\Sigma_{k+1 \vert k}^i = A_i \Sigma_{k\vert k}^i A^\top_i + W_i,\\
			\numberthis\label{eq: stochastic covariance at event-trigger}
			&\Sigma_{k\vert k }^i = \Sigma_{k \vert k-1}^i -{K}_k^i C_i \Sigma_{k \vert k-1}^i.
			\end{align*}
			The initial conditions are $\hat{x}_{0\vert -1}^i[1] = 0$ and $\Sigma_{0 \vert -1}^i = \Omega_0^i$.
		\end{subequations}
	\end{theorem}
	At last, we define $Z_k^i$ as the positive definite covariance matrix of the innovation signal in \eqref{eq: innovation at zero} i.e.
	\begin{align}
	\label{eq: innovation covariance}
	Z_k^i = \mathbb{E}\left[\zeta_k^i\left(\zeta_k^i\right)^\top\right] = K_{k}^i C_i \Sigma_{k\vert k-1}^i\left(K_{k}^i C_i\right)^\top + V_i.
	\end{align}
	\subsection{Cascade of Estimators}
	In this section we conclude the description of the network estimators by giving the state estimators for each decision maker $j = 2, \ldots, L+1$ in the chain $\mathcal{H}[i]$. For a system $i \in \{1, \ldots, N\}$, the optimal state estimator at decision maker $j \in \{2, \ldots, L+1\}$, i.e., $\mathcal{S}^i[2], \ldots, \mathcal{S}^i[L]$ and $\mathcal{C}^i$ is 
	\begin{subequations}
		\label{eq: cascade of estimators}
		\begin{align}
		\label{eq: estimations in the cascade}
		&\hat{x}_{k\vert k-1}^i[j] = A_i\hat{x}_{k-1\vert k-1}^i[j] + B_iu_{k - 1}^i\\
		&\hat{x}_{k\vert k}^i[j] = \begin{cases}
		\hat{x}_{k\vert k - d[j]}^i[j - 1], \quad & \text{if } \delta_{k-d[j]}^i[j-1] = 1,\\
		\hat{x}_{k\vert k-1}^i[j], & \text{otherwise}.
		\end{cases}
		\end{align}
	\end{subequations}
	where $\delta_{t}^i[j] \triangleq 0$ for all $t < 0$ and $\hat{x}_{0\vert -1}^i[j] = \hat{x}_{0\vert -1}^i[1]$. Moreover, the optimal state estimate at $j=1$ is by given in Theorem~\ref{thm: kalman filter}.
    {
	In what follows it will be useful to highlight the dependency of estimators in \eqref{eq: cascade of estimators} as function of Age of Information (AoI), i.e., difference between the current time and the time in which the most recent received measurement was generated (see Fig.~\ref{fig: age evolution}).
	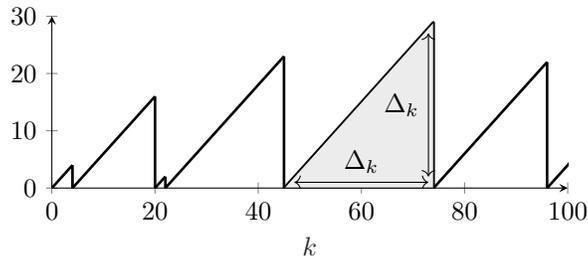
\begin{figure}[h!]
		\centering
		\begin{tikzpicture}
		
		\begin{axis}[%
		width=2.7in,
		height=0.9in,
		at={(3.0in,0.429in)},
		scale only axis,
		xmin=0,
		xmax=100,
		extra x ticks={},
		xlabel style={font=\color{white!15!black}},
		xlabel={$k$},
		ymin=0,
		ymax=30,
		axis background/.style={fill=white},
		axis x line=bottom,
		axis y line=left
		]
		\addplot [color=black, line width=1.0pt, forget plot]
		table[row sep=crcr]{%
			0	0\\
			4	4\\
		};
		\addplot [color=black, line width=1.0pt, forget plot]
		table[row sep=crcr]{%
			4	0\\
			4	4\\
		};
		\addplot [color=black, line width=1.0pt, forget plot]
		table[row sep=crcr]{%
			4	0\\
			4	-0\\
		};
		\addplot [color=black, line width=1.0pt, forget plot]
		table[row sep=crcr]{%
			4	0\\
			4	-0\\
		};
		\addplot [color=black, line width=1.0pt, forget plot]
		table[row sep=crcr]{%
			4	0\\
			20	16\\
		};
		\addplot [color=black, line width=1.0pt, forget plot]
		table[row sep=crcr]{%
			20	0\\
			20	16\\
		};
		\addplot [color=black, line width=1.0pt, forget plot]
		table[row sep=crcr]{%
			20	0\\
			22	2\\
		};
		\addplot [color=black, line width=1.0pt, forget plot]
		table[row sep=crcr]{%
			22	0\\
			22	2\\
		};
		\addplot [color=black, line width=1.0pt, forget plot]
		table[row sep=crcr]{%
			22	0\\
			45	23\\
		};
		\addplot [color=black, line width=1.0pt, forget plot]
		table[row sep=crcr]{%
			45	0\\
			45	23\\
		};
		\addplot [color=black, line width=1.0pt, forget plot]
		table[row sep=crcr]{%
			45	0\\
			74	29\\
		};
		\addplot [color=black, line width=1.0pt, forget plot]
		table[row sep=crcr]{%
			74	0\\
			74	29\\
		};
		\addplot [color=black, line width=1.0pt, forget plot]
		table[row sep=crcr]{%
			74	0\\
			96	22\\
		};
		\addplot [color=black, line width=1.0pt, forget plot]
		table[row sep=crcr]{%
			96	0\\
			96	22\\
		};
		\addplot [color=black, line width=1.0pt, forget plot]
		table[row sep=crcr]{%
			96	0\\
			104	8\\
		};
		\addplot [color=black, line width=1.0pt, forget plot]
		table[row sep=crcr]{%
			104	0\\
			104	8\\
		};
		\addplot[area legend, draw=black, fill=white!85!gray, forget plot]
		table[row sep=crcr] {%
			x	y\\
			45	0\\
			74	0\\
			74	29\\
		}--cycle;
		%\node[fill=white!85!black,scale=2,pin=135:{$(3,24)$}] at (axis cs:65,10) {$\Delta_k = \frac{Y_k^2}{2}$};
		\coordinate (x0) at (axis cs:47,1) {};
		\coordinate (x1) at (axis cs:73,1) {};
		\coordinate (y0) at (axis cs:73,2) {};
		\coordinate (y1) at (axis cs:73,27) {};
		%\node[scale=1.2] at (axis cs:65,10) {$\Delta_k$};
		\draw[<->] (x0) --node [above,midway] {$\Delta_k$} (x1);
		\draw[<->] (y0) --node [left,midway] {$\Delta_k$} (y1);
		\end{axis}
		\end{tikzpicture}%
		\caption{Age of information $\Delta_k$ as function of time $k$ and triggering instants for a network with no delays. The AoI grows linearly when no packet is received otherwise it assumes its minimum value.}
		\label{fig: age evolution}
	\end{figure}
	The scheduling algorithm we propose in this article will partially depend on the AoI $\Delta_k^i[j]$, $j = 1, \ldots, L+1$, at the decision makers, and the difference in AoI between two consecutive decision makers $j$ and $j+1$, i.e., the relative AoI defined as $\nabla\Delta_k^i[j] \triangleq \Delta_k^i[j+1] - \Delta_k^i[j]$.}
	The AoI $\Delta_k^i[j]$ at decision maker $j \in \mathcal{H}[i]$ with respect to system $i$ is given by
	\begin{align*}
		\numberthis\label{eq: AoI definition}
		\Delta_k^i[j] = \begin{cases}
		\Delta_{k-d[j]}^i[j-1] + d[j], & \text{if } \delta_{k-d[j]}^i[j-1] = 1\\
		\Delta_{k-1}^i[j] + 1, & \text{otherwise.}
		\end{cases}
	\end{align*} 
	with the natural extension $\Delta_t^i[j]=0$ for all $t \leq 0$. It is immediate to see that $\Delta_k^i[1]$ is identically zero since we assumed $d[1]=0$. Additionally, it holds
	\begin{align*}
		\numberthis\label{eq: increasing AoI}
		0 = \Delta_k^i[1] \leq \Delta_k^i[2] \leq \cdots \leq \Delta_k^i[L+1].
	\end{align*}
	In the absence of packet loss we conclude that the sequence of AoIs $\Delta_{0:k}^i[j]$ represent a sufficient statistic for sequence of triggering variables $\delta_{0:k-d[j]}^i[j-1]$ for $j \in \{2, \ldots, L+1\}$. It is then possible to rewrite the state estimation equations in \eqref{eq: cascade of estimators} as 
	\begin{align*}
	\hat{x}_{k\vert k}^i[j] = \hat{x}_{k \big\vert k-\Delta_k^i[j]}^i[j] =& A^{\Delta_k^i[j]}_i \left(\hat{x}_{k-\Delta_k^i[j] \big\vert k-\Delta_k^i[j] - 1}^i[1]  + \zeta_{k - \Delta_k^i[j]}^i\right) + \sum_{t = k - \Delta_k^i[j]}^{k - 1} A^{k - 1 -t}_i B_i u_t^i.
	\end{align*}
	
	\subsection{Error Dynamics}
	Highlighting $\Delta_k^i[j]$ in the expression of the state $x_k^i $ yields
	\begin{align*}
	x_k^i = & A^{\Delta_k^i[j]}_i x_{k-\Delta_k^i[j]}^i + \sum_{t = k - \Delta_k^i[j]}^{k - 1} A^{k - 1 -t}_i \left(B_i u_t^i + w_t^i\right).
	\end{align*}
	Let $ e_{k}^i[j]$ and $e_{k}^{i,-}[j]$ denote the estimation error and prediction error dynamics, respectively, i.e.
	\begin{subequations}
		\begin{align}
		\label{eq: definition estimation error}
		& e_{k}^i[j] \triangleq x_k^i - \hat{x}_{k\vert k}^i[j]\\
		\label{eq: definition prediction error}
		& e_{k}^{i,-}[j] \triangleq x_k^i - \hat{x}_{k\vert k - 1}^i[j]
		\end{align}
	\end{subequations}
	From our assumption $\Delta_k^i[1] = 0$ $\forall k$, we can rewrite \eqref{eq: definition estimation error} as 
	\begin{align*}
		e_{k}^i[j] &= x_{k}^i - \hat{x}_{k\vert k}^i[1] +\sum_{l = 1}^{j-1}\left( \hat{x}_{k\vert k-\Delta_k^i[l]}^i[l] - \hat{x}_{k\vert k - \Delta_k^i[l+1]}^i[l+1]\right)\\
		\numberthis\label{eq: estimation error as sum of mismatch errors}
		& = e_{k}^i[1] +\sum_{l = 1}^{j-1}\tilde{x}_{k \vert k}^i[l].
	\end{align*}
	We observe that $e_{k}^i[1]$ is the estimation error of a standard Kalman filter, as in Theorem~\ref{thm: kalman filter}, meanwhile $\tilde{x}_{k \vert k}^i[j]$ represents the mismatch error between decision maker $j$ and $j+1$ 
	\begin{align*}
		\numberthis\label{eq: mismatch error}
		\tilde{x}_{k \vert k}^i[j] & \triangleq \hat{x}_{k \vert k}^i[j] - \hat{x}_{k \vert k}^i[j+1]\\
		&= \hat{x}_{k\vert k-\Delta_k^i[l]}^i[j] - \hat{x}_{k\vert k - \Delta_k^i[l+1]}^i[j+1].
	\end{align*} 
	Recalling that $\hat{x}_{0\vert -1}^i[1] = 0$, the estimation error $e_{k}^i[1]$ evolves as
	\begin{align*}
	e_{k}^i[1] &= x_k^i - \hat{x}_{k \big\vert k}^i[1] \\
	&= A^k_i x_0^i + \sum_{t = 0}^{k-1} A_i^{k-1-t}\left(Bu_t^i + w_t^i\right) - A^k_i \hat{x}_{0\vert 0}^i[1] - \sum_{t = 0}^{k-1} A_i^{k-1-t}\left(Bu_t^i + \zeta_{t+1}^i\right)\\
	\numberthis\label{eq: standard Kalman filter error}
	& \overset{\eqref{eq: innovation at zero}}{=} A^k_i \left(x_0^i - \zeta_{0}^i\right) + \sum_{t = 0}^{k-1} A_i^{k-1-t}\left(w_t^i - \zeta_{t+1}^i\right).
	\end{align*}
	\subsection{Mismatch Error Dynamics and Relative AoI}
	\begin{proposition}
		\label{prop: mismatch error proposition}
		Let $\tilde{x}_{k \vert k}^i[j]$ be the mismatch error between DM $j$ and $j+1$ as defined in \eqref{eq: mismatch error}. Let $d[j]>0$ for all $j\in \{2, \ldots, L+1\}$. Then
		\begin{align*}
		\tilde{x}_{k \vert k}[j] = \sum_{t = k - \Delta_k^i[j] - \nabla\Delta_k^i[j]}^{k - \Delta_k^i[j] - 1} A^{k - 1 -t}_i \zeta_{t+1}^i.
		\end{align*}
		where $\nabla\Delta_k^i[j]$ represents the {relative AoI} and defined as
		\begin{align*}
			\numberthis\label{eq: inter aoi}
			\nabla\Delta_k^i[j] & \triangleq \Delta_k^i[j+1] - \Delta_k^i[j]\\
				 & = \delta_{k - d[j+1]}^i[j] \left(\Delta^i_{k+d[j+1]}[j] -\Delta_k^i[j] + d[j+1]\right) + \left(1 - \delta_{k - d[j+1]}^i[j]\right) \left(	\Delta_{k-1}^i[j+1] -\Delta_k^i[j] + 1  \right).
		\end{align*}
		Furthermore,
		\begin{subequations}
			\label{eq: statistical properties of error}
			\begin{align}
			\label{eq: covariance mismatch}
			& \mathbb{E}\left[\tilde{x}_{k \vert k}[j] \tilde{x}_{k \vert k}[j]^\top \right] = \mathbb{E}\left[\sum_{t = k - \Delta_k^i[j] - \nabla\Delta_k^i[j]}^{k - \Delta_k^i[j] - 1} A^{k - 1 -t}_i \zeta_{t+1}^i \left(A^{k - 1 -t}_i \zeta_{t+1}^i \right)^\top\right],\\
			\label{eq: independent mismatches}
			&\mathbb{E}\left[\tilde{x}_{k \vert k}[j] \tilde{x}_{k \vert k}[j^\prime]^\top \right] = 0, \text{ for } j \neq j^\prime,\\
			\label{eq: crosscorelation kalman error and mismatches}
			& \mathbb{E}\left[e_{k}^i[1] \tilde{x}_{k \vert k}[j]^\top \right] = -\mathbb{E}\left[\tilde{x}_{k \vert k}[j] \tilde{x}_{k \vert k}[j]^\top\right].
			\end{align}
		\end{subequations}
		%where $Z_{t}^i$ is the positive definite covariance matrix of the innovation given in \eqref{eq: innovation covariance}.
	\end{proposition}
	\begin{IEEEproof}
	 We note that the AoI at any decision maker is defined with respect to the time in which the most recent received measurement was generated at the sensor. It follows that
		\begin{align*}
		\hat{x}_{k \vert k}^i[j] & = \hat{x}_{k \vert k-\Delta_k^i[j]}^i[1]\\
		& = A^{\Delta_k^i[j]}_i \hat{x}_{k-\Delta_k^i[j] \big\vert k-\Delta_k^i[j]}^i[1] + \sum_{t = k - \Delta_k^i[j]}^{k - 1} A^{k - 1 -t}_i B_i u_t^i.
		\end{align*}
	Let $\Delta_k^i[j+1] - \Delta_k^i[j]=  \nabla\Delta_k^i[j]$ be the forward difference operator. We can write
	\begin{align*}
	\hat{x}_{k \vert k}^i[j+1] & = \hat{x}_{k \big\vert k-\Delta_k^i[j+1]}^i[1] = \hat{x}_{k \big\vert k-\Delta_k^i[j] - \nabla\Delta_k^i[j]}^i[1]\\
	& = A^{\Delta_k^i[j] + \nabla\Delta_k^i[j]}_i \hat{x}_{k-\Delta_k^i[j]-\nabla\Delta_k^i[j] \big\vert k-\Delta_k^i[j] - \nabla\Delta_k^i[j]}^i[1] + \sum_{t = k - \Delta_k^i[j] - \nabla\Delta_k^i[j]}^{k - 1} A^{k - 1 -t}_i B_i u_t^i.
	\end{align*}
	Highlighting $\Delta_k^i[j]+  \nabla\Delta_k^i[j]$ in the expression of $\hat{x}_{k \vert k}^i[j]$ we obtain 
	\begin{align*}
	\hat{x}_{k \vert k}^i[j] = & A^{\Delta_k^i[j] + \nabla\Delta_k^i[j]}_i \hat{x}_{k-\Delta_k^i[j]-\nabla\Delta_k^i[j] \big\vert k-\Delta_k^i[j] - \nabla\Delta_k^i[j]}^i[1]\\
	& + \sum_{t = k - \Delta_k^i[j] - \nabla\Delta_k^i[j]}^{k - \Delta_k^i[j] - 1} A^{k - 1 -t}_i \left(B_i u_t^i + \zeta_{t+1}^i\right) + \sum_{t = k - \Delta_k^i[j]}^{k - 1} A^{k - 1 - t}_i B_i u_t^i.
	\end{align*}
	From $\tilde{x}_{k \vert k}[j] = \hat{x}_{k \big\vert k-\Delta_k^i[j]}^i[j] - \hat{x}_{k \big\vert k-\Delta_k[j+1]}^i[j+1]$ it follows immediately that
	\begin{align*}
		\tilde{x}_{k \vert k}[j] =  \sum_{t = k - \Delta_k^i[j] - \nabla\Delta_k^i[j]}^{k - \Delta_k^i[j] - 1} A^{k - 1 -t}_i \zeta_{t+1}^i.
	\end{align*}
	From the previous equation the equality in \eqref{eq: covariance mismatch} is straight-forward.
	Furthermore, we observe that \eqref{eq: standard Kalman filter error} is composed of independent and uncorrelated terms. Equality \eqref{eq: crosscorelation kalman error and mismatches} is proven by observing that any cross-correlation term in
	$\mathbb{E}\left[e_{k}^i[1] \tilde{x}_{k \vert k}[j]^\top \right]$ will vanish.
	Finally, under the assumption that $d[j]>0$ for all $j\in \{2, \ldots, L+1\}$ the $j$-th decision maker in $\mathcal{H}[i]$ is knows $\color{red}\Delta_{0:k+d[j+1]-1}^i[j+1]$ at time $k$ since it is function of $\delta_{0:k-1}^i[j], \Delta_{0:k}^i[j]$. Additionally, we observe that $\tilde{x}_{k \vert k}[j]$ is function of the innovation $\zeta_{t}^i$ for $t \in \left[k-\Delta_{k}[j+1] + 1, k-\Delta_{k}[j]\right]$. From \eqref{eq: increasing AoI}, it follows that if $j \neq j^\prime$, $\tilde{x}_{k \vert k}[j]$ and $\tilde{x}_{k \vert k}[j^\prime]$ are function of innovations from disjoint intervals and therefore independent. That is $\mathbb{E}\left[\tilde{x}_{k \vert k}[j] \tilde{x}_{k \vert k}[j^\prime]^\top \right] = 0, \text{ for } j \neq j^\prime$. 
%	Therefore
%	\begin{align*}
%	& \mathbb{E}\left[\tilde{x}_{k \vert k}[j] \tilde{x}_{k \vert k}[j^\prime]^\top \right] = 0, \text{ for } j \neq j^\prime, \\
%	&\mathbb{E}\left[e_{k}^i[0]e_{k}^i[0]^\top\right] = \Sigma_{k \vert k}^i,\\
%	& \mathbb{E}\left[e_{k}^i[0] \tilde{x}_{k \vert k}[j]^\top \right] = \mathbb{E}\left[\tilde{x}_{k \vert k}[j] \tilde{x}_{k \vert k}[j]^\top \right].
%	\end{align*}
%	To conclude we observe that $\tilde{x}_{k \vert k}[j]$ is function of the innovation $\zeta_{t}^i$ for $t \in \left[k-\Delta_{k}[j+1] + 1, k-\Delta_{k}[j]\right]$. 
	\end{IEEEproof}
	
	\section{Optimal Certainty Equivalence Controller}
	{
	In \cite{ramesh2011} the authors showed that in a single-loop single-hop event-triggered networked control system dual effect in general exists. In addition, they showed that the certainty equivalence principle holds if and only if the triggering policy is independent of the control policy. Motivated by the work in \cite{molin2013} and \cite{ramesh2011}, we assume the scheduling policies to be function of the primitive random variables and innovation, i.e., independent of the control policies, and prove that the certainty equivalent controller is optimal in our general multi-loop multi-hop setting.}
	\begin{theorem}
		\label{thm: optimal u}
		Let Assumption~\ref{ass: broadcast of estimates instead of states} hold. Let the scheduling policies $\pi = \{\pi[1], \ldots, \pi[L]\}$ be a function of the innovation, the primitive random variables and the preceding scheduling variables, i.e.,
		\begin{align}
			\label{eq: triggering policies function of innovation}
			\delta_k^i[j] = \begin{cases}
			\pi_k^i[1]\left(\hat{x}_{0\vert -1}^i[1], \zeta_{0:k}^i, \delta_{0:k-1}^i[1] \right), & \text{if } j = 1\\
			\pi_k^i[j]\left(\hat{x}_{0\vert -1}^i[1], \zeta_{0:k-{D}[j]}^i, \delta_{0:k-{D}[j]}^i[1], \delta_{0:k-{D}[j] + d[1]}^i[2], \ldots, \delta_{0:k-d[j]}^i[j - 1], \delta_{0:k-1}^i[j]\right), & \text{otherwise}.
			\end{cases}
		\end{align}
		where ${D}[j]$ is the total delay at decision maker $j$ as given in \eqref{eq: total delay at j} in the decision chain $\mathcal{H}[i]$. If the triggering laws $\pi = \{\pi[1], \ldots, \pi[L]\}$ are given by \eqref{eq: triggering policies function of innovation} then the optimal control laws $\mu^\star$ minimizing $\sum_{i=1}^{N} J^i(\mu, \pi)$ of Problem~\ref{prob: optimization problem} are certainty equivalence controllers given by 
		\begin{equation}
		\label{eq: optimal controller}
		{u^i_k}^\star = -L_k^i \hat{x}_{k\vert k}^i[L+1],
		\end{equation}
		for $i \in \{1, \ldots, N\}$ and $\hat{x}_{k\vert k}^i[L+1]$ defined in \eqref{eq: cascade of estimators} and
		\begin{align*}
		&L_k^i = (R_k^i + B^\top_i S_{k+1}^{i} B_i)^{-1}B_i^\top S_{k+1}^i A_i,\\
		&S_k^i = Q_k^i + A^\top_i  \left(S_{k+1}^i - S_{k+1}^i B_i\left(R_i + B_i^\top S_{k+1}^i B_i\right)^{-1}B_i^\top S_{k+1}^i\right) A_i.
		\end{align*}
		with initial condition $S_{T} = \Lambda_i$. Moreover, defining $\Gamma_k^i \triangleq {L_k^i}^\top\left(R_k^i + B^\top_i S_{k+1}^i B_i\right)L_k^i$ the optimal cost for sub-system $i \in \{1, \ldots, N\}$ is
		\begin{align*}
		\numberthis \label{eq: optimal control cost}
		J^i(\mu^\star, \pi)	= & \mathbb{E} \left[{x_0^i}^\top S_0^ix_0^i\right] + \mathbb{E} \left[\sum_{k = 0}^{T-1} {w_k^i}^\top S_{k+1}^iw_k^i\right]  \\
		& + \mathbb{E} \left[\sum_{k=0}^{T-1}\left(x_k^i - \hat{x}_{k\vert k}^i[L]\right)^\top \Gamma_k^i\left(x_k^i - \hat{x}_{k\vert k}^i[L]\right) \right].
		\end{align*}
	\end{theorem}
	\begin{IEEEproof}
		We begin the proof by observing that the triggering policies defined in \eqref{eq: triggering policies function of innovation} can also be written only as function primitive random variable and preceding triggering policies. For a given system $i \in \{1, \ldots, N\}$ we can rewrite the triggering policy of its first scheduler as
		\begin{align*}
		\delta_0^i[1] =& 
		\pi_0^i[1]\left(\hat{x}_{0\vert -1}^i[1], \zeta_{0}^i\right),\\
		\delta_1^i[1] =& 
		\pi_1^i[1]\left(\hat{x}_{0\vert -1}^i[1], \zeta_{0:1}^i, \delta_0^i[1]\right) = \pi_1^i[1]\left(\hat{x}_{0\vert -1}^i[1], \zeta_{0:1}^i, \pi_0^i[1]\left(\hat{x}_{0\vert -1}^i[1], \zeta_{0}^i\right)\right),\\
		\delta_2^i[1] =& 
		\pi_2^i[1]\left(\hat{x}_{0\vert -1}^i[1], \zeta_{0:2}^i, \delta_{0}^i[1], \delta_{1}^i[1]\right)\\
		= & \pi_1^i[1]\left(\hat{x}_{0\vert -1}^i[1], \zeta_{0:2}^i, \pi_0^i[1]\left(\hat{x}_{0\vert -1}^i[1], \zeta_{0}^i\right), \pi_1^i[1]\left(\hat{x}_{0\vert -1}^i[1], \zeta_{0:1}^i, \pi_0^i[1]\left(\hat{x}_{0\vert -1}^i[1], \zeta_{0}^i\right)\right)\right),\\
		\vdots. &
		\end{align*}
		It is easily concluded that the sequence $\delta_{0:k}^i[1]$ is independent of the control law for any $k \geq 0$. In general, it can be seen by continuous substitution that seen that $\delta_{0:k}^i[j]$ is independent of the control law for any $j\in \{1, \ldots, L\}$ and $i \in \{1, \ldots, N\}$. As a consequence, the choice of the scheduling policies $\pi$ determines uniquely whether the constraints in Problem~\ref{prob: optimization problem} are satisfied. Therefore solving the minimization Problem~\ref{prob: optimization problem} for the given scheduling policies is equivalent to minimizing the unconstrained problem $\sum_{i = 1}^{N} J^i(\mu, \pi)$. The resulting objective function is a purely quadratic and decoupled, i.e., minimizing $\sum_{i = 1}^{N} J^i(\mu, \pi)$ over all admissible control policies reduces to minimizing a purely quadratic function $J^i(\mu, \pi)$ for each system since no physical coupling is present. 
		Additionally, we can show that since $\delta_{0:k}^i[j]$ is independent of the control law for the given scheduling policies in \eqref{eq: triggering policies function of innovation} so is the estimation error $x_k - \mathbb{E}\left[x_k^i \vert \mathcal{I}_k^i\right]$ at the controller (for proof see \cite{molin2013}).
		Finally, from standard stochastic control theory \cite{bertsekas1995dynamic}, it can be proven that the control law \eqref{eq: optimal controller} is optimal for fixed scheduling policies.
		%For system $i$ and a scheduler $j$, when a scheduling policy satisfying \eqref{eq: triggering policies function of innovation} exists, the random decision variable $\delta_k^i[j]$ is described by a function of primitive random variables and \textit{delayed} innovation that is independent of the choice of the control policy. It implies that the choice of the scheduling policy determines uniquely whether the constraints in Problem~\ref{prob: optimization problem} are satisfied irrespectively from the control policy. Thus, solving the optimization problems for a fixed scheduling policy reduces to minimizing over all admissible control laws. The resulting objective function is purely quadratic, and tools from stochastic control can be applied \cite{bertsekas1995dynamic} and \cite[Lemma~1]{molin2013}.
	\end{IEEEproof}
	\begin{remark}
		In the remainder of this article we will assume that the scheduling policies are independent of the control actions and function of primitive random variable as in \eqref{eq: triggering policies function of innovation}. As pointed out in \cite{ramesh2011}, since the scheduling policies \eqref{eq: triggering policies function of innovation} are independent of the past control actions, the controller has no dual effect. 
		It should be noted that the resulting closed-loop system is not optimal in general since a controller with dual
		effect may result in lower cost \cite{ramesh2011, witsenhausen1986}.
	\end{remark}
	\section{Distributed Value of Information-Based Scheduling}
	In this section we introduce the distributed value of information (dVoI) metric. Informally speaking, the dVoI is a metric based on locally available information of each scheduler and it represents the quality of data between any two neighboring decision makers. In the remainder of this article we prove that dVoI-based scheduling policies guarantee a certain level of performance for both control and communication. 
	\subsection{Problem Decomposition}
	In this paragraph we will decomposition the global optimization problem in Problem~\ref{prob: optimization problem} into independent by introducing a multi-hop dynamic programming algorithm. As first step, we observe that the constraints in \eqref{eq: triggering cost} are soft-constraints. The existence of a Lagrangian multiplier for a single-hop is guaranteed under mild conditions \cite{molin2014, soleymani2018value}. However, the results also apply to the multi-hop setting as shown in the following. It follows that Problem~\ref{prob: optimization problem} can be rewritten as
	\begin{align*}
	\numberthis\label{eq: dual problem}
	\underset{\pi^i}{\operatorname{inf}} \ J^i({\mu^i}^\star, \pi^i) + \sum_{j = 1}^{L}\lambda[j] r^i[j], \qquad i = 1, \ldots, N.
	\end{align*}
	where the Lagrangian multiplier $\lambda[j]$ is with respect to the coupled constraints \eqref{eq: triggering cost}, and the optimal control policies are given in Theorem~\ref{thm: optimal u}.
%	{\color{red}
%	Without loss of generality in what follows we will assume that $d[j]=1,$ for $j\neq 1$. The case in which $d[j]>1$ can be dealt with by introducing dummy nodes with optimal scheduling policy $\delta_k^i[j^\prime]=1$ for a given dummy node $j^\prime$ and for all $k, i$. We will provide an example after the Theorem~\ref{thm: VoI-based scheduling}.}
	
	Without loss of generality in what follows we will assume that $d[j]=1,$ for $j\neq 1$. The methodology presented also holds for the case $d[j]>1$ but with some additional notation difficulty.
	\begin{lemma}
	\label{lem: hop DP}
	Consider the set of schedulers $j \in \mathcal{H}[i]\backslash\mathcal{C}^i$, i.e., $j\in \{1, \ldots, L\}$ for a given system $i \in \{1, \ldots, N\}$. Let $\lambda[j]$ for $j \in \mathcal{H}[i]\backslash\mathcal{C}^i$, i.e., $j\in \{1, \ldots, L\}$,, be the optimal Lagrangian multipliers corresponding to the set of constraints in \eqref{eq: triggering cost}. Moreover, let $d[j]=1$ for $j > 1$. Under the admissible scheduling policies \eqref{eq: triggering policies function of innovation}, solving Problem~\ref{prob: optimization problem} is equivalent to solving the Bellman-like equation
	\begin{align}
	\label{eq: value function of scheduler}
	\mathbb{H}_{\pi, k}^i[j] = \underset{\delta^i_{k}[j]}{\operatorname{min}} \ \mathbb{E} \left[\mathbb{G}_k^i[j] + \mathbb{H}_{\pi, k+1}^i[j+1]\Bigg\vert \mathcal{J}^i_k[j]\right],
	\end{align}
	with initial conditions $\mathbb{H}_{\pi, t}^i[L+1]=0$ for all $t$, and $\mathbb{H}_{\pi, T}^i[l]=0$ for all $l$. Moreover, $\mathbb{G}_k^i[j]$ represents a hop-cost-to-go given by
	\begin{align*}
	\numberthis\label{eq: hop cost}
	\mathbb{G}_k^i[j] =& \sum_{l=j}^{L}  {-2 e_{k+1\vert k+1}^i[1]}^\top \Gamma_{k+1}^i \tilde{x}_{k+1\vert k+1}^i[l]+ {\tilde{x}_{k+1\vert k+1}^i[l]}^\top \Gamma_{k+1}^i \tilde{x}_{k+1\vert k+1}^i[l]  +  \lambda[j]\delta_k^i[j].
	\end{align*}
\end{lemma}
\begin{IEEEproof}
	Observing that, under the assumption that $d[j]=1$ for $j \neq 1$, each scheduler $j\in \{1, \ldots, L\}$ is actually agnostic of any other scheduler $j^\prime$ such that $j^\prime < j-1$, the scheduling policy in \eqref{eq: triggering policies function of innovation} can then be rewritten as
	\begin{align*}
	\delta_k^i[j] = \begin{cases}
	\pi_k^i[1]\left(\hat{x}_{0\vert -1}^i[1], \zeta_{0:k}^i , \delta_{0:k-1}^i[1] \right), & \text{if } j = 1\\
	\pi_k^i[j]\left(\hat{x}_{0\vert -1}^i[1], \zeta_{0:k-{D}[j]}^i, \delta_{0:k-1}^i[j - 1], \delta_{0:k-1}^i[j]  \right), & \text{otherwise}.
	\end{cases}
	\end{align*}
	It follows that the team of schedulers $\{1, \ldots, L\}$ is a sequential team and we can therefore apply dynamic programming techniques. Consider the situation at time $k$ and at scheduler $j$. The information set $\mathcal{J}_{k}[j]$ has been observed and the problem is to determine the scheduling strategy $\pi^i_k[j]$ such that the cost function \eqref{eq: dual problem} is minimal.
	We can rewrite the cost as
	\begin{align*}
	& \mathbb{E} \left[\sum_{k=0}^{T-1}\left(x_k^i - \hat{x}_{k\vert k}^i[L+1]\right)^\top \Gamma_k^i\left(x_k^i - \hat{x}_{k\vert k}^i[L+1]\right) \right]  \overset{\eqref{eq: estimation error as sum of mismatch errors}}{=} \mathbb{E} \left[\sum_{k=0}^{T-1}\left(e_{k\vert k}^i[1] +\sum_{j = 1}^{L} \tilde{x}_{k\vert k}^i[j]\right)^\top \Gamma_k^i\left(e_{k\vert k}^i[1] +\sum_{j = 1}^{L} \tilde{x}_{k\vert k}^i[j]\right) \right]\\
	& \overset{\eqref{eq: statistical properties of error}}{=} \mathbb{E}\left[\sum_{k=0}^{T-1} {e_{k\vert k}^i[1]}^\top \Gamma_k^ie_{k\vert k}^i[1] +  \sum_{j=1}^{L} {-2 e_{k\vert k}^i[1]}^\top \Gamma_k^i \tilde{x}_{k\vert k}^i[j]+ {\tilde{x}_{k\vert k}^i[j]}^\top \Gamma_k^i \tilde{x}_{k\vert k}^i[j]\right].
	\end{align*}
	Since the term $ {e_{k\vert k}^i[1]}^\top \Gamma_k^ie_{k\vert k}^i[1]$ is independent of all scheduling policies, we will ignore it in the subsequent analysis.	We observe that 
	\begin{align*}
	%\numberthis\label{eq: fetching out j}
	& \mathbb{E} \left[\sum_{l=1}^{L} \sum_{k=0}^{T-1}  {-2 e_{k\vert k}^i[1]}^\top \Gamma_k^i \tilde{x}_{k\vert k}^i[l]+ {\tilde{x}_{k\vert k}^i[l]}^\top \Gamma_k^i \tilde{x}_{k\vert k}^i[l] +  \lambda[l]\delta_k^i[l]\right] = \\
	& \mathbb{E} \left[\sum_{l=1}^{j-1} \sum_{k=0}^{T-1}  {-2 e_{k\vert k}^i[1]}^\top \Gamma_k^i \tilde{x}_{k\vert k}^i[l]+ {\tilde{x}_{k\vert k}^i[l]}^\top \Gamma_k^i \tilde{x}_{k\vert k}^i[l] +  \lambda[l]\delta_k^i[l]\right] + \mathbb{E} \left[\sum_{l=j}^{L} \sum_{k=0}^{T-1}  {-2 e_{k\vert k}^i[1]}^\top \Gamma_k^i \tilde{x}_{k\vert k}^i[l]+ {\tilde{x}_{k\vert k}^i[l]}^\top \Gamma_k^i \tilde{x}_{k\vert k}^i[l] +  \lambda[l]\delta_k^i[l]\right].
	\end{align*}
	Only the last term depends on scheduling variables $\delta_{0:T-1}^i[j]$. Moreover, from \eqref{eq: cascade of estimators} and \eqref{eq: mismatch error} $ \tilde{x}_{k\vert k}^i[l]$ depends on $\delta_{k-1}^i[l]$. Assuming that a minimum exists, it follows that
	\begin{align*}
	\underset{\delta^i_{k}[j]}{\operatorname{min}} \ & \mathbb{E} \left[\sum_{l=j}^{L} \sum_{t=k}^{T-1}  {-2 e_{t+1\vert t+1}^i[1]}^\top \Gamma_{t+1}^i \tilde{x}_{t+1\vert t+1}^i[l]+ {\tilde{x}_{t+1\vert t+1}^i[l]}^\top \Gamma_{t+1}^i \tilde{x}_{t+1\vert t+1}^i[l] +  \lambda[j]\delta_t^i[j] \right] \\ & = \mathbb{E} \left[ \underset{\delta^i_{k}[j]}{\operatorname{min}} \ \mathbb{E} \left[\sum_{l=j}^{L} \sum_{t=k}^{T-1}  {-2 e_{t+1\vert t+1}^i[1]}^\top \Gamma_{t+1}^i \tilde{x}_{t+1\vert t+1}^i[l]+ {\tilde{x}_{t+1\vert t+1}^i[l]}^\top \Gamma_{t+1}^i \tilde{x}_{t+1\vert t+1}^i[l] +  \lambda[j]\delta_t^i[j] \Bigg\vert \mathcal{J}^i_k[j] \right] \right],
	\end{align*}
	where the outer expected value is with respect to the distribution of $\mathcal{J}^i_k[j]$, and the minimum is taken with respect to the admissible policies \eqref{eq: triggering policies function of innovation}. Repeating the argument for $k+1, \ldots, T-1$ under the assumption that all minima exist and are unique, we obtain
	\begin{align*}
	\underset{\delta^i_{k:T-1}[j]}{\operatorname{min}} \ & \mathbb{E} \left[\sum_{l=j}^{L} \sum_{t=k}^{T-1}  {-2 e_{t+1\vert t+1}^i[1]}^\top \Gamma_{t+1}^i \tilde{x}_{t+1\vert t+1}^i[l]+ {\tilde{x}_{t+1\vert t+1}^i[l]}^\top \Gamma_{t+1}^i \tilde{x}_{t+1\vert t+1}^i[l] +  \lambda[j]\delta_t^i[j] \right] = \mathbb{E} \left[\mathbb{V}_{k}^i[j] \right],
	\end{align*}
	where the minima are taken with respect to the admissible policies \eqref{eq: triggering policies function of innovation} and where the function $V_{k}^i[j]$ satisfies the following Bellman equation 
	\begin{align*}
	\numberthis\label{eq: V value function}
	\mathbb{V}_{k}^i[j] = \underset{\delta^i_{k}[j]}{\operatorname{min}} \ & \mathbb{E} \left[\sum_{l=j}^{L}  {-2 e_{k+1\vert k+1}^i[1]}^\top \Gamma_{k+1}^i \tilde{x}_{k+1\vert k+1}^i[l]+ {\tilde{x}_{k+1\vert k+1}^i[l]}^\top \Gamma_{k+1}^i \tilde{x}_{k+1\vert k+1}^i[l] +  \lambda[j]\delta_k^i[j] +\mathbb{V}_{k+1}^i[j]\Bigg\vert \mathcal{J}^i_k[j] \right],
	\end{align*}
	with $\mathbb{V}_{T}^i[j]=0$. Analogously, repeating the argument for schedulers $j+1, \ldots, L$ under the assumption that all minima exist and are unique, we obtain
	\begin{align*}
	\underset{\delta^i_{k}[j], \ldots, \delta^i_{k}[L]}{\operatorname{min}} \ & \mathbb{E} \left[\sum_{l=j}^{L} \sum_{t=k}^{T-1}  {-2 e_{t+1\vert t+1}^i[1]}^\top \Gamma_{t+1}^i \tilde{x}_{t+1\vert t+1}^i[l]+ {\tilde{x}_{t+1\vert t+1}^i[l]}^\top \Gamma_{t+1}^i \tilde{x}_{t+1\vert t+1}^i[l] +  \lambda[j]\delta_t^i[j] \right] = \mathbb{E} \left[\mathbb{W}_{k}^i[j] \right],
	\end{align*}
	where the minima are taken with respect to the admissible policies \eqref{eq: triggering policies function of innovation} and where the function $\mathbb{V}_{k}^i[j]$ satisfies the following Bellman equation 
	\begin{align*}
	\numberthis\label{eq: W value function}
	\mathbb{W}_{k}^i[j] = \underset{\delta^i_{k}[j]}{\operatorname{min}} \ & \mathbb{E} \left[\sum_{t=k}^{T-1}  {-2 e_{t+1\vert t+1}^i[1]}^\top \Gamma_{t+1}^i \tilde{x}_{t+1\vert t+1}^i[j]+ {\tilde{x}_{t+1\vert t+1}^i[j]}^\top \Gamma_{t+1}^i \tilde{x}_{t+1\vert t+1}^i[j] +  \lambda[j]\delta_t^i[j]+ \mathbb{W}_{k}^i[j+1]\Bigg\vert \mathcal{J}^i_k[j]\right],
	\end{align*}
	with $\mathbb{W}_{k}^i[L]=0$.
	Finally, by putting \eqref{eq: V value function} and \eqref{eq: W value function} together we obtain the Bellman-like equation
	\begin{align*}
	\mathbb{H}_{\pi, k}^i[j] = \underset{\delta^i_{k}[j]}{\operatorname{min}} \ & \mathbb{E} \left[\sum_{t=k}^{T-1}  {-2 e_{t+1\vert t+1}^i[1]}^\top \Gamma_{t+1}^i \tilde{x}_{t+1\vert t+1}^i[j]+ {\tilde{x}_{t+1\vert t+1}^i[j]}^\top \Gamma_{t+1}^i \tilde{x}_{t+1\vert t+1}^i[j] +  \lambda[j]\delta_k^i[j] \right.\\
	& \left. + \sum_{l=j+1}^{L}  {-2 e_{k+1\vert k+1}^i[1]}^\top \Gamma_{k+1}^i \tilde{x}_{k+1\vert k+1}^i[l]+ {\tilde{x}_{k+1\vert k+1}^i[l]}^\top \Gamma_{k+1}^i \tilde{x}_{k+1\vert k+1}^i[l] +  \mathbb{H}_{\pi, k+1}^i[j+1]\Bigg\vert \mathcal{J}^i_k[j]\right],
	\end{align*}
	with initial conditions $\mathbb{H}_{\pi, t}^i[L+1]=0$ for all $t$, and $\mathbb{H}_{\pi, T}^i[l]=0$ for all $l$. The conclusion follows noticing that $\tilde{x}_{t+1 \vert t+1}^i[j]$ is independent of $\delta_{k}^i[j]$ for $t>k$.
\end{IEEEproof}
\begin{remark}
	\label{rem: hard hop DP}
	It follows from Lemma~\ref{lem: hop DP} that the optimal policy of each scheduler $j \in \mathcal{H}[i]\backslash\mathcal{C}^i$, i.e., $j\in \{1, \ldots, L\}$, depends also on the optimal policies of the schedulers $j^\prime \in \{j+1, \ldots, L\}$. Due to this coupling the problem is therefore hard to solve. Furthermore, since the characterization of the value functions is not easily obtainable the problem of finding optimal scheduling policies is still open. 
\end{remark}

{In the following we propose sub-optimal policies for the problem in \eqref{eq: dual problem} which outperforms the standard periodic policy ${\bar{\pi}^i[j]}$ characterized by  $\delta_k^i[j] = 1$ for all $k \in \{0, \ldots, T-1\}$, $j \in \{1, \ldots, L\}$, and $1 \in \{1, \ldots, N\}$.} Formally, under the assumptions of Lemma~\ref{lem: hop DP}, Problem~\ref{prob: optimization problem} can be written as
\begin{align*}
J({\mu}^\star, \pi) = \sum_{i=1}^{N} J^i({\mu^i}^\star, \pi^i) + \sum_{j = 1}^{L}\lambda[j] r^i[j].
\end{align*}
We are interested in sub-optimal policies $\pi^\star$ such that
\begin{align*}
\underset{\pi}{\min} \ J({\mu}^\star, \pi) \leq J({\mu}^\star, \pi^\star) \leq J({\mu}^\star, \bar{\pi}).
\end{align*}
That is, the sub-optimal policy $\pi^\star$ outperforms the periodic policy  $\bar{\pi}$ in the sense that it attains a lower cost.
The following theorem characterizes the sub-optimal policy $\pi^\star$.
\begin{lemma}
	\label{lem: decomposition}
	Consider the set of schedulers $j \in \mathcal{H}[i]\backslash\mathcal{C}^i$, i.e., $j\in \{1, \ldots, L\}$ for a given system $i \in \{1, \ldots, N\}$. Let $\lambda[j]$ for $j\in \{1, \ldots, L\}$, be the optimal Lagrangian multipliers corresponding to the set of constraints in \eqref{eq: triggering cost}, and let $d[j]=1$ for $j \neq 1$. Furthermore, let ${\bar{\pi}^i[j]}$ be a periodic policy with  $\delta_k^i[j] = 1$ for all $0 \leq k \leq T-1$. For each scheduler consider the Bellman-like equation in  \eqref{eq: value function of scheduler}. The set of policies $\left\lbrace \pi^{i\star}[j], \bar{\pi}^i[j+1], \ldots, \bar{\pi}^i[L] \right\rbrace$ outperforms the set of periodic policies $\left\lbrace \bar{\pi}^i[j], \bar{\pi}^i[j+1], \ldots, \bar{\pi}^i[L] \right\rbrace$ if the periodic policy $\bar{\pi}^i[j]$ is replaced by the sub-optimal policy $\pi^{i\star}[j]$.
\end{lemma}
\begin{IEEEproof}
	We prove this by inductions. Clearly, $0=\mathbb{H}_{\pi^\star, T}^i[L+1] \leq \mathbb{H}_{\bar{\pi}, T}^i[L+1] = 0$. Assume that the claim holds at time $k+1$ and scheduler $j+1$. We have
	\begin{align*}
	\mathbb{H}_{\pi^\star, k}^i[j] & = \underset{\delta^i_{k}[j]}{\operatorname{min}} \ \mathbb{E} \left[\mathbb{G}_k^i[j] + \mathbb{H}_{\pi, k+1}^i[j+1]\Bigg\vert \mathcal{J}^i_k[j]\right] \\
	& = \mathbb{E} \left[\mathbb{G}_k^i[j] + \mathbb{H}_{\pi, k+1}^i[j+1]\Bigg\vert \mathcal{J}^i_k[j], \delta_k^{i\star}[j]\right]\\
	& \leq  \mathbb{E} \left[\mathbb{G}_k^i[j] + \mathbb{H}_{\bar{\pi}, k+1}^i[j+1]\Bigg\vert \mathcal{J}^i_k[j], \delta_k^{i\star}[j]\right]\\
	& \leq \mathbb{H}_{\bar{\pi}, k}^i[j],
	\end{align*}
	where the first inequality comes from the induction hypothesis and the second inequality from the definition of the sub-optimal scheduling policy $\bar{\pi}^i[j]$.
\end{IEEEproof}
\subsection{Distributed Value Of Information}
{In this paragraph we finally introduce the distributed value of information (dVoI) metric.  As the cost metric \eqref{eq: dual problem} is the aggregation of local control and communication objective functions, the proposed dVoI-based scheduling policy must guarantee a certain level of performance for both control and communication. In fact, as given in the following theorem, the dVoI-based scheduling policy represents a suboptimal and distributed solution with guaranteed performance to the trade-off problem \eqref{eq: dual problem} in a multi-loop multi-hop networked control systems.}
\begin{theorem}[Distributed VoI-based Scheduling]
	\label{thm: VoI-based scheduling}
	Consider the set of schedulers $j \in \mathcal{H}[i]\backslash\mathcal{C}^i$, i.e., $j\in \{1, \ldots, L\}$ for a given system $i \in \{1, \ldots, N\}$. Let $\lambda[j]$ for $j \in \mathcal{H}[i]\backslash\mathcal{C}^i$, i.e., $j\in \{1, \ldots, L\}$, be the optimal Lagrangian multipliers corresponding to the set of constraints in \eqref{eq: triggering cost}. Moreover, let $d[j]=1$ for $j > 1$. For each scheduler $j$ consider the optimization problem~\eqref{eq: dual problem}. The periodic policy ${\bar{\pi}^i[j]}$, with  $\delta_k^i[j] = 1$ for all $0 \leq k \leq T-1$, is outperformed by the VoI-based sub-optimal policy given by 
	\begin{align}
	\label{eq: optimal scheduling policy}
	{\delta_k^i[j]}^\star = 
	\begin{cases}
	1, \quad \text{if } \textbf{dVoI}_{k}^i[j]<0,\\
	0, \quad \text{otherwise}.
	\end{cases}
	\end{align}
	if $k\leq T-(L+1-j)$ and ${\delta_k^i[j]}^\star=0$ otherwise. The Value of Information $\textbf{dVoI}_{k}^i[j]$, defined as the gain in the cost when a measurement is successfully sent as	opposed to when it is blocked, is given by
	\begin{align*}
	\numberthis\label{eq: value of information}
	\textbf{dVoI}_{k}^i[j] &= \lambda[j]-\left(\tilde{x}_{k \vert k}^i[j]\right)^\top \left({A_i^{L+1-j}}\right)^\top \Gamma^i_{ k+(L+1-j)} \left(A_i^{L+1-j}\right)\tilde{x}_{k \vert k}^i[j].
	\end{align*}
\end{theorem}
\begin{remark}
	The optimal closed-loop scheduling policy is a value-of-information-based policy that depends particularly on the estimation innovation and the mismatch estimation errors at the schedulers. The first set of \textbf{dVoI}s, i.e., $\textbf{dVoI}^i_k[1]$ for $i=1, \ldots, N$, behaves qualitatively as the VoI in \cite{soleymani2018value}: it is mainly function of the estimation error. In fact, in the special case of a single-hop the dVoI-based policy is equivalent to the well-known threshold-based scheduling policy where the mismatch error is used as the state of the scheduler and a fixed $\lambda$ as the transmission cost \cite{soleymani2018value}, \cite{molin2014}, \cite{klgel2019joint}. Formally, for $L = 1$, it follows that
\begin{align*}
{\delta_k^i}^\star = \begin{cases}
1, \quad \text{if } \left(\tilde{x}_{k \vert k}^i\right)^\top A_i^\top \Gamma^i_{ k+1} A_i\tilde{x}_{k \vert k}^i > \lambda, \\
0, \quad \text{otherwise}.
\end{cases}
\end{align*}
	where the mismatch error is $\tilde{x}_{k \vert k}^i = \hat{x}_{k \vert k}^i - \hat{x}_{k \vert k-\Delta_k^i}^i$, with $\Delta_k^i\triangleq \Delta_k^i[2]$. 
	
	The remaining schedulers $j \ne 1$ are such that the  $\textbf{dVoI}^i_k[j]$, for $i=1, \ldots, N$,  are mainly function of locally available information, i.e., AoI, relative AoI with respect to its successor, eigenvalues of the system and of the noise covariances.
\end{remark}
\begin{remark}
	From the definition of Value of Information in \eqref{eq: value of information} we notice that it is a function of the mismatch error $\tilde{x}_{k \vert k}^i[j]$ given in 		\eqref{eq: mismatch error}
	\begin{align*}
		\tilde{x}_{k \vert k}^i[j] = \hat{x}_{k \vert k}^i[j] - \hat{x}_{k \vert k}^i[j+1] = \hat{x}_{k\vert k-\Delta_k^i[j]}^i[1] - \hat{x}_{k\vert k - \Delta_k^i[j+1]}^i[1].
	\end{align*}
	In the definition of the estimators in \eqref{eq: cascade of estimators} we assumed $u_{0:k-1}^i$ to be available to all decision makers. However, from the previous equality we can conclude that in order for a scheduler $j>1$ to compute its value of information it needs to know its age of information, the relative age of information with respect to its successor, and its position in the decision chain $\mathcal{H}[i]$ defined in \eqref{eq: chain DMs}. That is, it is function of locally available information. Furthermore, the value-of-information-based policy depends particularly on the estimation innovation and the mismatch estimation errors at the schedulers and it is therefore a determining metric in networked control systems contrarily to the age of information metric. Moreover, the set of first schedulers $\mathcal{S}[1]$ will have a value on information strongly related to what the authors in \cite{soleymani2018value} defined as value of information. The subsequent schedulers, $\mathcal{S}[2], \ldots, \mathcal{S}[L]$, however, are mainly function of locally available information such as the age of information, the relative age of information with respect to their successor, eigenvalues of the system, and some delayed state estimates. 
\end{remark}
\begin{remark}
	As a final remark to this theorem, we highlight that at each scheduler the computation of the $N$ utilities of information, i.e., $\textbf{dVoI}_{k}^i[j]$ for $i=1, \ldots, N$, is done independently between sub-systems, and between hops.	 Additionally, the the definition is consistent through all schedulers and sub-systems as defined in \eqref{eq: value of information}. Computing the $\textbf{dVoI}_{k}^i[j]$ at each time incurs a computational complexity of $O(n_i^3)$, with $n_i$ being the dimension of the state of sub-system $i$. The overall complexity for scheduler $j$ is then $O(N*n_i^3)$.  In case of a big horizon $T$, if the gains $\Gamma_k^i$, $i=1, \ldots,N$ converge and can be substituted with $\Gamma^i$, then computational complexity can be reduced to $O(N*n_i^2)$ if the products $\left({A_i^{L+1-j}}\right)^\top \Gamma^i \left(A_i^{L+1-j}\right)$ are computed once and then stored. 
\end{remark}
\begin{remark}
	
	As discussed in Remark~\ref{rem: different number of hops}, the result of Theorem~\ref{thm: VoI-based scheduling} applies in a straight-forward manner to the case where the control loops $i=1,\ldots, N$ are closed through a different number of communication hops $L_i$. As an example, in the special case of $N=2$, and number of hops $L_1 \neq L_2$, with $L_1 < L2$, we have that Theorem~\ref{thm: VoI-based scheduling} yields  
		\begin{align*}
			\textbf{dVoI}_{k}^1[j] &= \lambda[j]-\left(\tilde{x}_{k \vert k}^1[j]\right)^\top \left({A_1^{L_1+1-j}}\right)^\top \Gamma^1_{ k+(L_1+1-j)} \left(A_1^{L_1+1-j}\right)\tilde{x}_{k \vert k}^1[j], & j = 1, \ldots, L_1,\\
			\textbf{dVoI}_{k}^2[j] &= \lambda[j]-\left(\tilde{x}_{k \vert k}^2[j]\right)^\top \left({A_2^{L_2+1-j}}\right)^\top \Gamma^2_{ k+(L_2+1-j)} \left(A_2^{L_2+1-j}\right)\tilde{x}_{k \vert k}^2[j], & j = 1, \ldots, L_2.
	\end{align*}
	The Lagrangian multipliers $\lambda[j]$, for $j = 1, \ldots, L_1$, governs the communication rate of the two control loops meanwhile $\lambda[j]$, for $j = L_1, \ldots, L_2$, governs the communication rate of the second control loop.
\end{remark}
\section{Information Constraints In Practice}
	In the derivation of the optimal estimators in Section~\ref{sec: network estimators}, we assumed that the history of the control actions $u_{0:k-1}^i$ is available to every scheduler. In the trivial case of a single-hop, i.e., $L=1$, it is a reasonable assumption since the communication channel is assumed ideal and the scheduler knows the information available at the controller. In a general multi-hop setting, however, it is a strong assumption since it requires a one-step delay feedback channel from the controller to every scheduler independent of the number of hops $L$. We will assume that a given scheduler $j$ can compute the control actions $u_{0:k-(L+1-j)}^i$. This can be achieved by redefining the information set in \eqref{eq: information set schedulers} as
	\begin{equation}
	\label{eq: new information set}
	\tilde{\mathcal{J}}_{k}^i[j] \triangleq \begin{cases} \mathcal{J}_k^i[L], & \text{if } j = L, \\
	\mathcal{J}_k^i[j] \cup \left\lbrace \delta_{0:k-(L+1-l)}^i[l], l=j+1, \ldots, L\right\rbrace, & \text{otherwise}.
	\end{cases} 
	\end{equation} 
	{In other words, the communication acknowledgment channel from decision maker $j+1$ to $j$ will require exactly $N(L-j)$ bits in order for scheduler $j$ to be able to infer $u_{0:k-(L+1-j)}^i$, for all $i=1, \ldots, N$.} We observe that the case of $j=L$, no feedback channel is necessary since we assume an ideal channel communication channel. %The design methodology proposed hereafter, still holds in case a feedback channel is not present. However, the trade-off problem will generally perform  worse.
	Under the new information sets \eqref{eq: new information set}, at time $k$ the first scheduler has access to the history of control actions until time $k-L$, i.e. $u_{0:k-L}^i$.{As a consequence, the Kalman filter in Theorem~\ref{thm: kalman filter} is optimal only until time step $k-L+1$, i.e. $\hat{x}_{k-L+1\vert k-L+1}[1]$, since the most recent history $u_{k-L+1:k-1}^i$ is not available to the first scheduler.} It is therefore necessary to design an additional estimator for the time steps $k-L+2:k$.
	\subsection{\textbf{dVoI} with Unknown Inputs} 
	\label{subsec: unknow inputs}
	Several works in literature \cite{KITANIDIS1987775, DAROUACH1997717, GILLIJNS2007111, janczak2006state} propose the joint design of unbiased minimum-variance input and state estimator. In what follows we extend the work in \cite{janczak2006state} and define an optimality criterion to ensure minimal error of the state estimates under the influence of unknown inputs. We will therefore consider the prediction equation in \eqref{eq: kalman filter equations at event-trigger} as a dynamical system for which optimal control actions $u_{k-L+1:k-1}^i$ should be calculated. Formally,
	\begin{align*}
	J_{\text{estimator}}= \sum_{t=k-L+2}^{k} \left(y_t^i - C_i\hat{x}_{t\vert t-1}^i[1]\right)^\top Q_{i,t}^e \left(y_t^i - C_i\hat{x}_{t\vert t-1}^i[1]\right) + (u_{t-1}^i)^\top R_{i,t-1}^e u_{t-1}^i,
	\end{align*}
	where $Q_{i,t}^e$ and $R_{i,t-1}^e$, for $t=k-L+2,\ldots, k$ are positive definite matrices determining the weights of the corresponding errors and control estimates. Therefore, solving 
	\begin{align*}
		\numberthis \label{eq: estimation cost}
		\underset{u_{k-L+1:k-1}^i}{\operatorname{min}} \ 	J_{\text{estimator}}
	\end{align*}
	corresponds to finding the optimal control estimates $\hat{u}_{k-L+1:k-1}^i$. The solution to \eqref{eq: estimation cost} can be calculated through Bellman dynamic programming \cite{bertsekas1995dynamic, janczak2006state}.
In general, the covariance of the control estimation error will influence the value of information. However, for tractability issue and using Theorem~\ref{thm: VoI-based scheduling} we approximate the distributed value of information as
\begin{align*}
\numberthis\label{eq: approximated dVoI}
\textbf{dVoI}_{k}^i[j] \simeq  \lambda[j] - \mathbb{E} \left[\left(\tilde{x}_{k \vert k}^i[j]\right)^\top \left({A_i^{L+1-j}}\right)^\top \Gamma^i_{ k+(L+1-j)} A_i^{L+1-j}\tilde{x}_{k \vert k}^i[j]\Big\vert \mathcal{J}^i_k[j]\right].
\end{align*}	
From the previous equality it is evident that that from the point of view of scheduler $j$, $j$ and $j+1$ have the same belief of the control estimates. Therefore, $\hat{u}_{k-L+1:k-1}^i$ will not affect $\textbf{dVoI}_{k}^i[j]$. Using this fact, it is sufficient for the first scheduler in the chain  $\mathcal{H}[i]$ to compute control estimates, subtract the influence of ${u}_{0:k-L}^i$ and $\hat{u}_{k-L+1:k-1}^i$ from the state estimate, forward the control-free state estimate to successive decision maker in $\mathcal{H}[i]$ based on the value of information. {In conclusion, using the approximation \eqref{eq: approximated dVoI}, the dVoI-based scheduling policies can therefore be easily computed in practice in a multi-loop multi-hop networked control system.}
\section{Numerical Example}
In this section, we show an application of the theoretical framework we developed in this article. As previously stated, numerous heterogeneous systems coupled through a shared communication network may be evaluated with no added effort. However, for the sake of simplicity, we will focus on a single-loop system. 

\begin{figure}[ht!]
	\centering
	\includegraphics[width=0.5\textwidth]{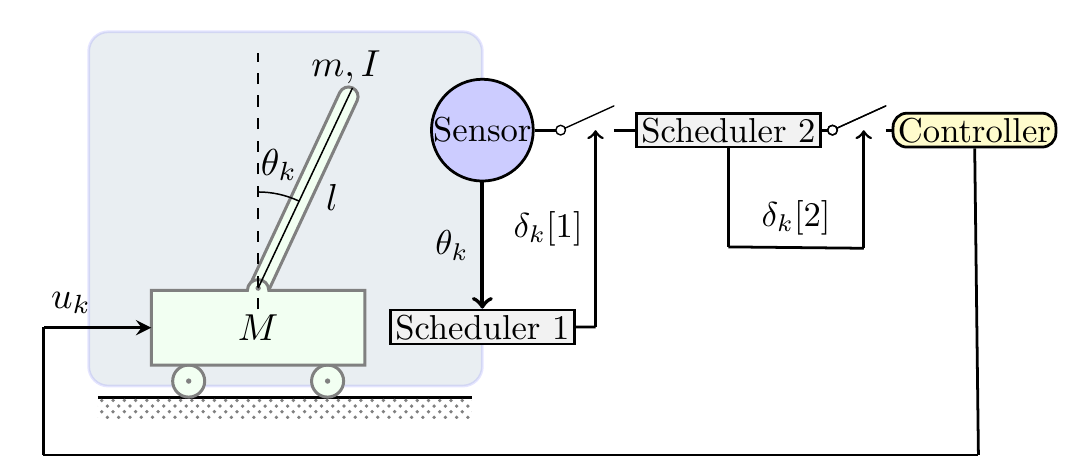}
	\caption{Model of an inverted pendulum on a cart. The sensor is insourced and controller is outsourced. The sensor is connected to the controller through a two-hop communication network.}
	\label{fig: cart 2 hop}
\end{figure}

Consider an inverted pendulum on a cart, in Fig.~\ref{fig: cart 2 hop}, observed by an internal sensor, where the sensor is connected to the controller through a two-hop communication network.
Similarly to \cite{soleymani2018value}, we assume the following parameters: $I =6*10^{-3}$ kg m$^2$, $m =0.2$ kg, $l =0.6$ m, $g =9.81$ m/s${^2}$, cart mass $M =0.5$ kg, and $b =0.1$ N/m/sec. 
The sensor can only measure the position and the pitch angle. The time horizon is $T=200$, noise covariances are given by
\begin{align*}
	V = 10^{-3} * \begin{bmatrix}
	2 & 0 \\ 0 &1
	\end{bmatrix}, \quad  & 
	W = 10^{-4} * \begin{bmatrix}
	6 & 3 & 1 & 6\\ 
	3 & 8 & 3 & 4\\ 
	1 & 3 & 7 & 6\\ 
	6 & 4 & 6 & 31
	\end{bmatrix}.
\end{align*} 
We assume a sampling time of $100$ Hz, initial condition $x_0 = [0, 0, 0.2, 0]$, $\Omega_0 = W$. %Furthermore, we assume that the controller is located two-hops away from the sensor, i.e., $L=2$. 
The cost function \eqref{eq: dual problem} is specified with weighting matrices $\Lambda = Q= \operatorname{blkdiag}\{1, 1, 1000, 1\}$, $R = 1$, and communication costs $\lambda[1] = 15$, $\lambda[2]= 30$. From Theorem~\ref{thm: optimal u} and Theorem~\ref{thm: VoI-based scheduling} under our assumptions the optimal control policy is $u_k = -L_k \hat{x}_{k|k}[3]$ given in \eqref{eq: optimal controller}, and the optimal scheduling policy is dVoI-based. According to our discussion in Subsection~\ref{subsec: unknow inputs}, for a two-hop network $u_{k-1}$ is unknown to the first scheduler at time $k$ and must be estimated using \eqref{eq: estimation cost}. Applying standard dynamic programming to \eqref{eq: estimation cost}, the estimate of  $u_{k-1}$ is given by
\begin{align*}
	\hat{u}_{k-1} =\left(B^\top C^\top Q^e C B + R^e\right)^{-1} B^\top C^\top Q^e \left(y_k - C A \hat{x}_{k-1|k-1}[1]\right),
\end{align*} 
where the weighting matrices are chosen as $Q^e = 100 I$ and $R^e=0.1$. The corresponding state estimate is
\begin{align*}
\hat{x}_{k \vert k}[1] = & A_i\hat{x}_{k -1\vert k-1} +\left(B-KCB\right)\left(B^\top C^\top Q^e C B + R^e\right)^{-1} B^\top C^\top Q^e \left(y_k - C A \hat{x}_{k-1|k-1}[1]\right)\\
& + K_k\left(y_k - C A \hat{x}_{k-1|k-1}[1]\right).
\end{align*}

\begin{figure*}[ht!]
\begin{minipage}{1\textwidth}
		\centering
		\begin{minipage}{.49\textwidth}
			\includegraphics[width=1\textwidth]{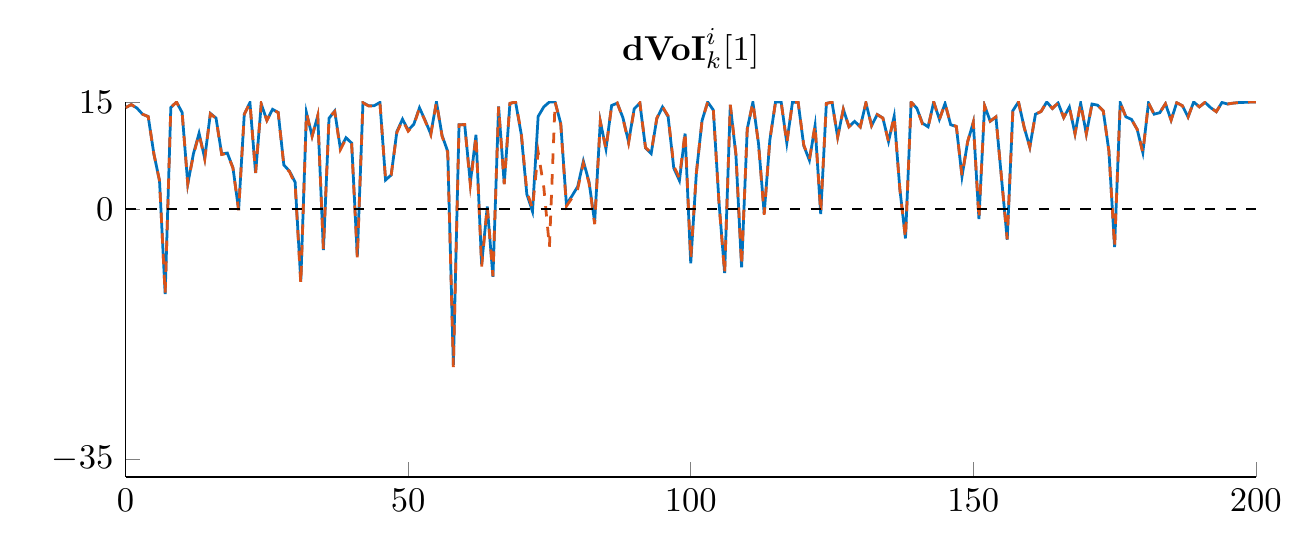}
		\end{minipage}
		\centering
		\begin{minipage}{.49\textwidth}
			\includegraphics[width=1\textwidth]{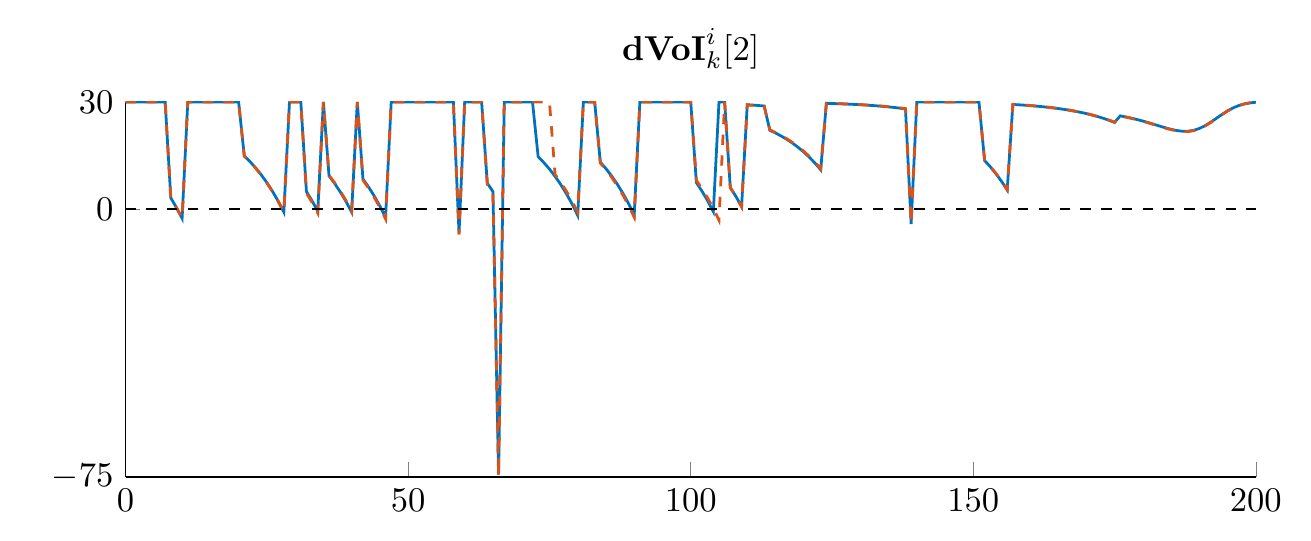}
		\end{minipage}
\end{minipage}
\begin{minipage}{1\textwidth}
	\centering
	\begin{minipage}{.49\textwidth}
		\includegraphics[width=1\textwidth]{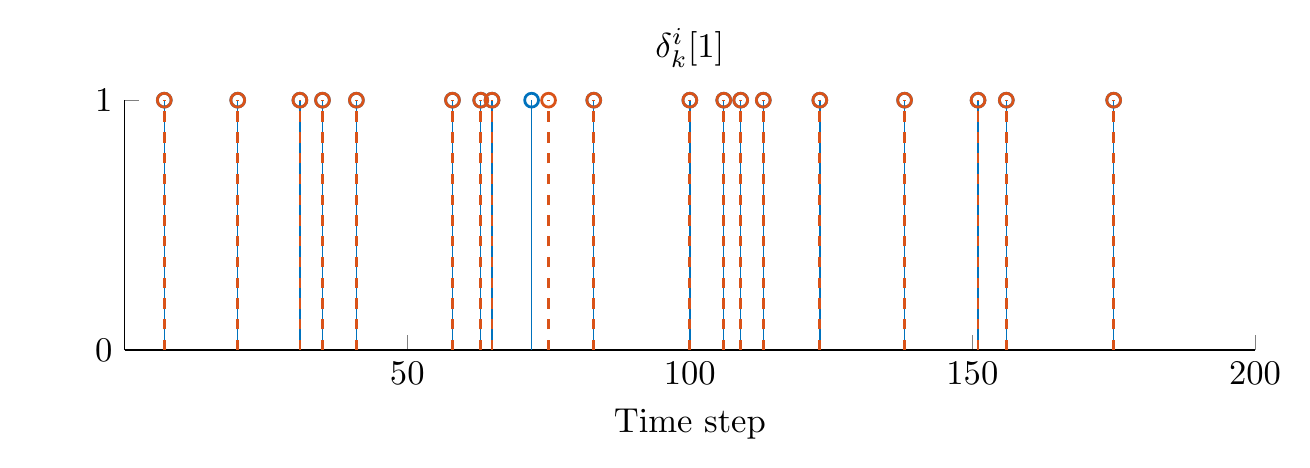}
		\captionof{figure}{VoI and triggering instants at Scheduler $1$. The solid lines represent the evolution when $u_{0:k-1}$ is assumed to be known at all schedulers. The dashed lines represent the evolution when $u_{k-1}$ is unknown to the first scheduler and estimated by solving \eqref{eq: estimation cost}.}
		\label{fig: scheduler 1 VoI}
	\end{minipage}
	\centering
	\begin{minipage}{.49\textwidth}
		\includegraphics[width=1\textwidth]{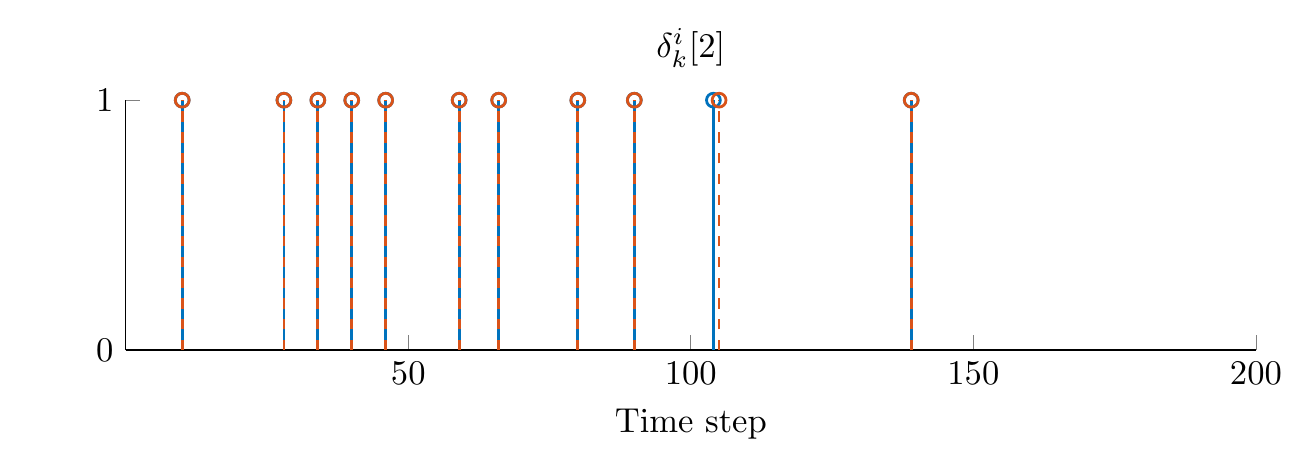}
		\captionof{figure}{VoI and triggering instants at Scheduler $2$. The solid lines represent the evolution when $u_{0:k-1}$ is assumed to be known at all schedulers. The dashed lines represent the evolution when $u_{k-1}$ is unknown to the first scheduler and estimated by solving \eqref{eq: estimation cost}.}
		\label{fig: scheduler 2 VoI}
	\end{minipage}
\end{minipage}
\begin{minipage}{1\textwidth}
	\begin{minipage}{.49\textwidth}
		{\includegraphics[width=1\textwidth]{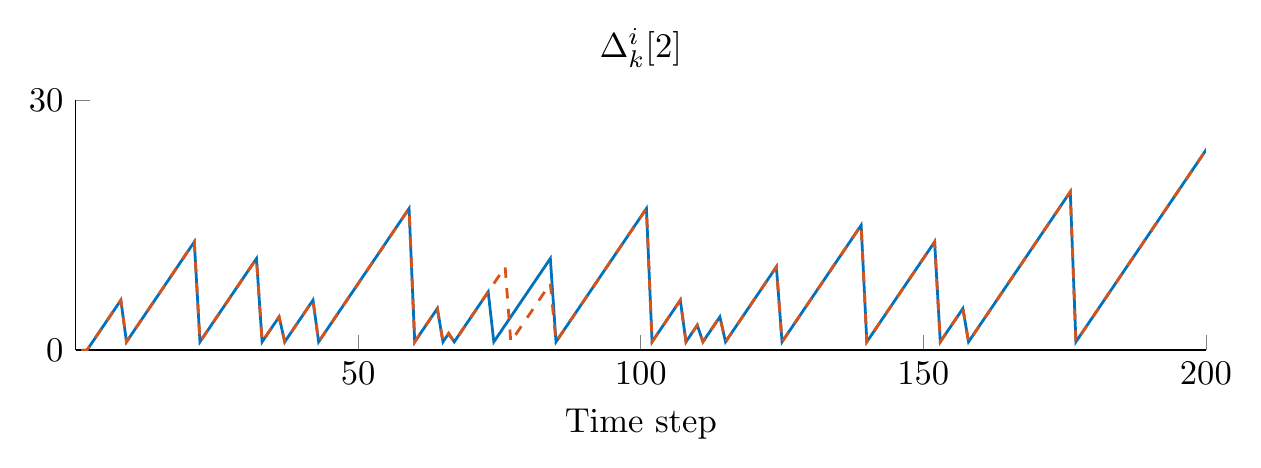}}
		\captionof{figure}{Age of information at scheduler $2$. The solid line represents the evolution when $u_{0:k-1}$ is assumed to be known at all schedulers. The dashed line represents the evolution when $u_{k-1}$ is unknown to the first scheduler and estimated by solving \eqref{eq: estimation cost}. As expected, the AoI drops to its minimum when an event is triggered.}
		\label{fig: AoI 2}
	\end{minipage}
	\begin{minipage}{.49\textwidth}
		{\includegraphics[width=1\textwidth]{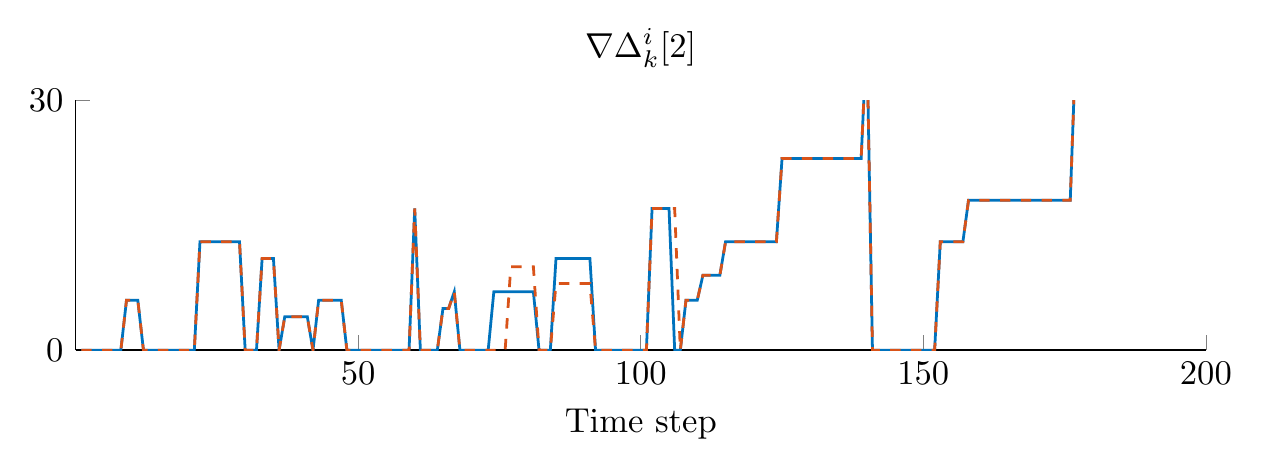}}
		\captionof{figure}{Relative age of information at scheduler $2$. The solid line represents the evolution when $u_{0:k-1}$ is assumed to be known at all schedulers. The dashed line represents the evolution when $u_{k-1}$ is unknown to the first scheduler and estimated by solving \eqref{eq: estimation cost}. When the rAoI is constant there no new measurements and therefore no new event. When the rAoI increases, although there is a new measurement the decrease in \textbf{dVoI} is not sufficient to trigger a new event.}
		\label{fig: relative AoI 2}
	\end{minipage}
\end{minipage}
\end{figure*}

In Fig~\ref{fig: scheduler 1 VoI} and \ref{fig: scheduler 2 VoI} we see the distributed value of information and the corresponding triggering instants.  As discussed earlier, we observe that the first \textbf{dVoI} is similar to the notion of VoI \cite{soleymani2018value}, i.e., it is mainly function of the system disturbances. The \textbf{dVoI} of the second scheduler, however, is mainly function of locally available information, i.e., AoI, relative AoI with respect to its successor, eigenvalues of the system and of the noise covariances. Depending on the Lagrangian multipliers, also the \textbf{dVoI} of the second scheduler could heavily depend on the system disturbances. 

From Fig~\ref{fig: AoI 2} and Fig~\ref{fig: relative AoI 2} we can fully infer the AoI at both schedulers and the controller since the AoI at the first scheduler is identically zero, meanwhile the AoI at the controller is the sum of the AoI at the second scheduler and the relative AoI between the second scheduler and the controller. That is, $\Delta_{k}[1]=0$ since $d[1]=0$, $\Delta_{k}[2]$ given in Fig~\ref{fig: AoI 2}, and $\Delta_{k}[3]=\Delta_{k}[2] + \nabla\Delta_{k}[2]$.  %Additionally, we observe that the instants in which the relative AoI of the second scheduler is zero, its VoI assumes its maximum value $\lambda[2]=30$. 
Furthermore, we observe that when the second scheduler and the controller have the same AoI, or if the decrease in \textbf{dVoI} is not large enough, then the optimal policy for the scheduler is to not transmit. We may conclude that AoI and rAoI are insufficient for control, which is consistent with the findings in \cite{ayan2019}.

\begin{figure}[h!]
	\begin{minipage}{1\textwidth}
	\begin{minipage}{.49\textwidth}
		\includegraphics[width=1\textwidth]{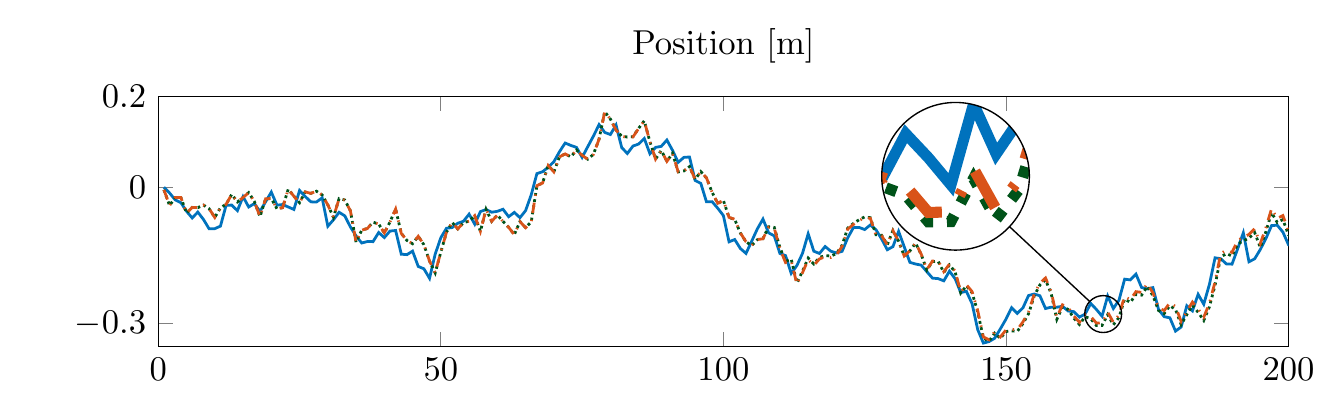}
	\end{minipage}
	\begin{minipage}{.49\textwidth}
	\includegraphics[width=1\textwidth]{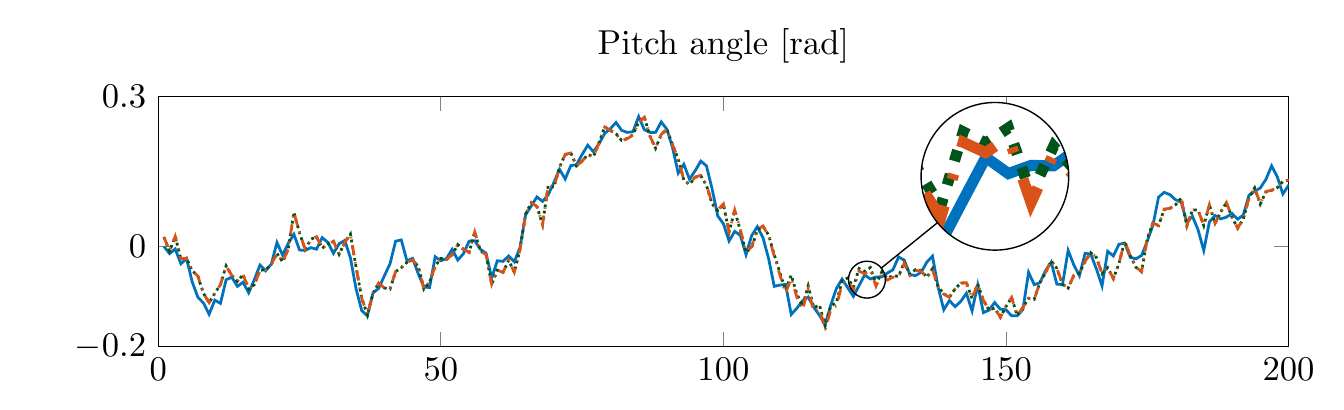}
	\end{minipage}
	\end{minipage}
	\begin{minipage}{1\textwidth}
	\begin{minipage}{.49\textwidth}
		{\includegraphics[width=1\textwidth]{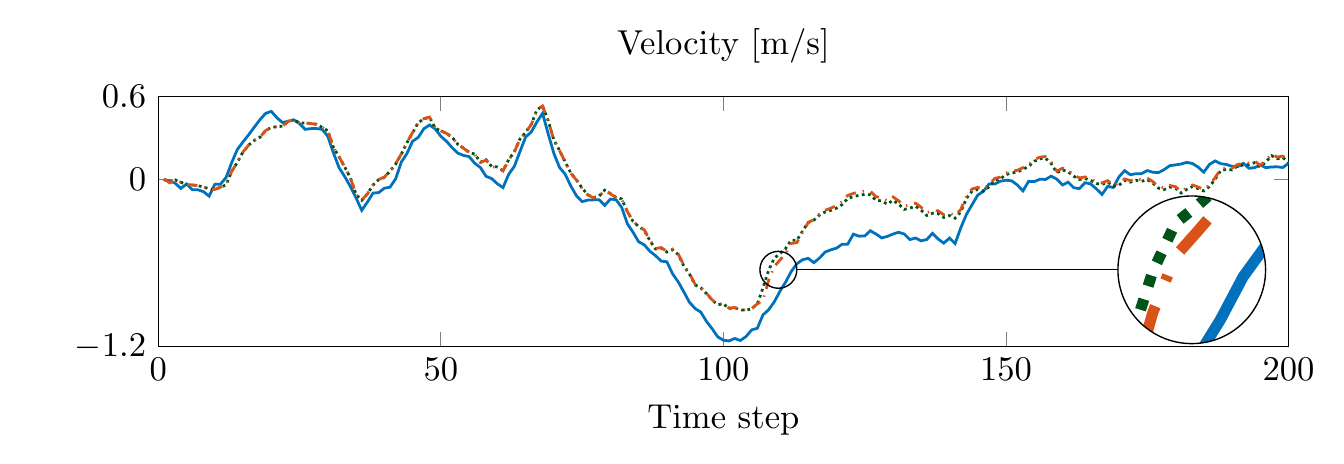}}
		\captionof{figure}{Position and velocity at the controller. The solid line represents the real state. The dotted line represents the state estimate when the history $u_{0:k-1}$ is assumed to be known at all schedulers.The dashdotted line represents the state estimate the history $u_{0:k-2}$ is assumed to be known at all schedulers and $u_{k-1}$ is unknown to the first scheduler and estimated by solving \eqref{eq: estimation cost}.}
		\label{fig: Position and velocity at the controller}
	\end{minipage}
	\begin{minipage}{.49\textwidth}
		{\includegraphics[width=1\textwidth]{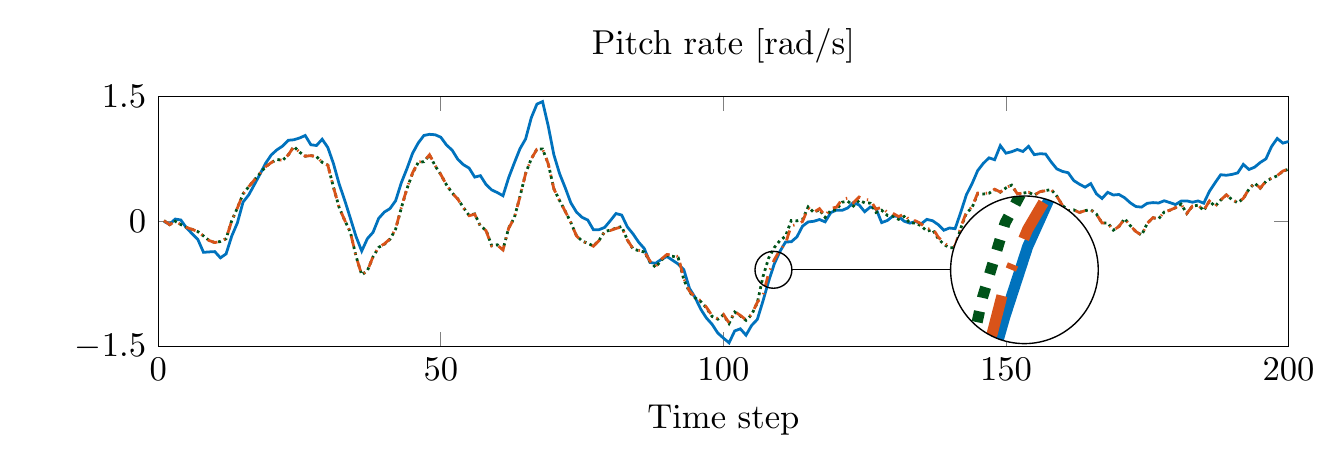}}	
		\captionof{figure}{Pitch angle and pitch rate at the controller. The solid line represents the real state. The dotted line represents the state estimate when the history $u_{0:k-1}$ is assumed to be known at all schedulers.The dashdotted line represents the state estimate the history $u_{0:k-2}$ is assumed to be known at all schedulers and $u_{k-1}$ is unknown to the first scheduler and estimated by solving \eqref{eq: estimation cost}.}
		\label{fig: Pitch angle and pitch rate at the controller}
	\end{minipage}
\end{minipage}
\end{figure}

\begin{figure}[h!]
	\centering
		{\includegraphics[width=0.5\textwidth]{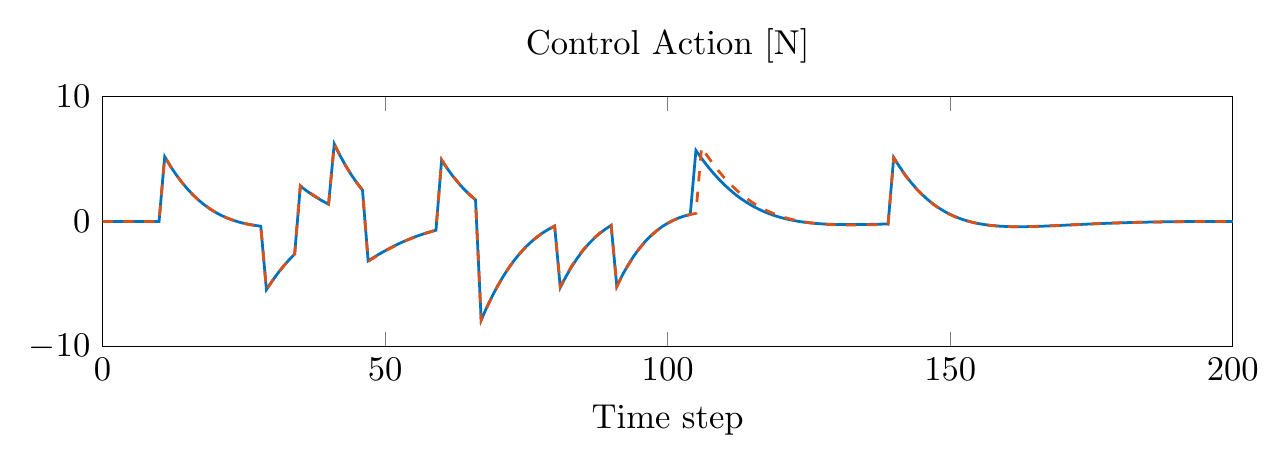}}
		\captionof{figure}{Signal applied by the controller. The solid line represents the evolution of the control signal $u_k$ when the history $u_{0:k-1}$ is assumed to be known at all schedulers. The dashed line represents the evolution of the control signal $u_k$ when the history $u_{0:k-2}$ is assumed to be known at all schedulers and $u_{k-1}$ is unknown to the first scheduler and estimated by solving \eqref{eq: estimation cost}.}
		\label{fig: control actions}
\end{figure}

{In Fig~\ref{fig: Position and velocity at the controller}-Fig~\ref{fig: control actions} we can see the states, estimates and the control actions when the whole control history is available to all the schedulers as opposed to when the last control action $u_{k-1}$ is not available at the first scheduler at is estimated instead.} In both scenario, the \textbf{dVoI} of the first and last scheduler went below zero $19$ and $11$ times, respectively.  Even though the communication rate was decreased by $90.5\%$ and $94.5\%$, for the first and last scheduler, respectively, the controller was still able to achieved a good overall performance.

%\begin{figure*}[ht!]
%	\centering
%	\begin{minipage}{.49\textwidth}
%		{\includegraphics[width=1\textwidth]{matlab/images/UoI1.pdf}}
%	\end{minipage}
%	\centering
%	\begin{minipage}{.49\textwidth}
%		{\includegraphics[width=1\textwidth]{matlab/images/UoI2.pdf}}
%	\end{minipage}
%\end{figure*}
\section{Conclusion}
In this article we provided a theoretical framework for the analysis and co-design of control and communication in a multi-loop multi-hop networked control systems. We reformulated the initial joint optimization problem as an equivalent Bellman-like equation and successively decomposed it into independent and distributed sub-problems of the same form. Furthermore, we proved that, under certain standard assumptions, the optimal closed-loop control policy is a certainty-equivalence policy and that the optimal closed-loop scheduling policy is a value-of-information-based policy that depends particularly on the estimation innovation and the mismatch estimation errors at the schedulers. At last, we discussed the role of the information constraints in a multi-hop communication channel, and how its influence on the proposed VoI-based policy can be minimized.
\bibliographystyle{IEEEtran}
\bibliography{References/refs}

\appendix
\section*{Proof of Theorem~\ref{thm: VoI-based scheduling}}
%\begin{IEEEproof}
	\subsubsection*{Part 1}
	First we compute the expression for the mismatch error $\tilde{x}_{k+1 \vert k+1}^i[j]$. From \eqref{eq: cascade of estimators} and equation \eqref{eq: mismatch error} we obtain
	\begin{align*}
	\tilde{x}_{k+1 \vert k+1}^i[j] & =  \hat{x}_{k+1 \vert k+1}^i[j] -  \hat{x}_{k+1 \vert k+1}^i[j+1] \\
	& =\hat{x}_{k+1 \vert k+1}^i[j] - \delta_k^i[j] \hat{x}_{k+1 \vert k}^i[j] - \left(1-\delta_k^i[j]\right)\hat{x}_{k+1 \vert k}^i[j+1]\\
	&= \delta_k^i[j]\left(\hat{x}_{k+1 \vert k+1}^i[j] - \hat{x}_{k+1 \vert k}^i[j]\right) 	+ \left(1-\delta_k^i[j]\right) \left(\hat{x}_{k+1 \vert k+1}^i[j] - \hat{x}_{k+1 \vert k}^i[j+1]\right).
	\end{align*}
	For $j>1$, \eqref{eq: cascade of estimators} it follows
	\begin{align*}
	\hat{x}_{k+1 \vert k+1}^i[j] &= \delta_{k}^i[j-1]\hat{x}_{k+1 \vert k}^i[j-1] + \left(1-\delta_{k}^i[j-1]\right)\hat{x}_{k+1 \vert k}^i[j]
	\end{align*}
	Therefore
	\begin{subequations}
		\label{eq: mismatch error espansion}
		\begin{align*}
		\tilde{x}_{k+1 \vert k+1}^i[j]  = & \ \delta_k^i[j]\left(\hat{x}_{k+1 \vert k+1}^i[j] - \hat{x}_{k+1 \vert k}^i[j]\right) 	+ \left(1-\delta_k^i[j]\right) \left(\hat{x}_{k+1 \vert k+1}^i[j] - \hat{x}_{k+1 \vert k}^i[j+1]\right)\\
		=& \ \delta_k^i[j]\left(\delta_{k}^i[j-1]\hat{x}_{k+1 \vert k}^i[j-1] + \left(1-\delta_{k}^i[j-1]\right)\hat{x}_{k+1 \vert k}^i[j] - \hat{x}_{k+1 \vert k}^i[j]\right)\\
		& + \left(1-\delta_k^i[j]\right) \left(\delta_{k}^i[j-1]\hat{x}_{k+1 \vert k}^i[j-1] + \left(1-\delta_{k}^i[j-1]\right)\hat{x}_{k+1 \vert k}^i[j] - \hat{x}_{k+1 \vert k}^i[j+1]\right) \\
		=&\ \delta_k^i[j]\left(\delta_{k}^i[j-1]\hat{x}_{k+1 \vert k}^i[j-1] - \delta_{k}^i[j-1]\hat{x}_{k+1 \vert k}^i[j]\right)\\
		& +\left(1-\delta_k^i[j]\right) \delta_{k}^i[j-1]\left(\hat{x}_{k+1 \vert k}^i[j-1] - \hat{x}_{k+1 \vert k}^i[j] + \hat{x}_{k+1 \vert k}^i[j] - \hat{x}_{k+1 \vert k}^i[j+1]\right) \\
		& + \left(1-\delta_k^i[j]\right)\left(1-\delta_{k}^i[j-1]\right)\left(\hat{x}_{k+1 \vert k}^i[j] - \hat{x}_{k+1 \vert k}^i[j+1]\right)\\
		= & \ \delta_k^i[j]\delta_{k}^i[j-1]A_i \tilde{x}_{k \vert k}^i[j-1] + \left(1-\delta_k^i[j]\right)\delta_{k}^i[j-1]A_i \tilde{x}_{k \vert k}^i[j-1]\\
		& + \left(1-\delta_k^i[j]\right)\delta_{k}^i[j-1]A_i \tilde{x}_{k \vert k}^i[j] + \left(1-\delta_k^i[j]\right)\left(1-\delta_{k}^i[j-1]\right)A_i\tilde{x}_{k \vert k}^i[j]\\
		\numberthis\label{eq: mismatch error espansion j>1}
		=&\ \delta_{k}^i[j-1]A_i \tilde{x}_{k \vert k}^i[j-1] + \left(1-\delta_k^i[j]\right)A_i\tilde{x}_{k \vert k}^i[j]
		\end{align*}
		Meanwhile for $j=1$, it follows that
		\begin{align*}
		\tilde{x}_{k+1 \vert k+1}^i[j]  = & \ \delta_k^i[j]\left(\hat{x}_{k+1 \vert k+1}^i[j] - \hat{x}_{k+1 \vert k}^i[j]\right) 	+ \left(1-\delta_k^i[j]\right) \left(\hat{x}_{k+1 \vert k+1}^i[j] - \hat{x}_{k+1 \vert k}^i[j+1]\right)\\
		=& \ \delta_k^i[j]\left(\hat{x}_{k+1 \vert k}^i[j] + \zeta_{k+1}^i - \hat{x}_{k+1 \vert k}^i[j]\right) 	+ \left(1-\delta_k^i[j]\right) \left(\hat{x}_{k+1 \vert k}^i[j] + \zeta_{k+1}^i - \hat{x}_{k+1 \vert k}^i[j+1]\right)\\
		\numberthis\label{eq: mismatch error espansion j=1}
		=& \ \zeta_{k+1}^i	+ \left(1-\delta_k^i[j]\right) A_i\tilde{x}_{k \vert k}^i[j].
		\end{align*}
	\end{subequations}
	We now look at the expression for 
	\begin{align*}
	\underset{\delta^i_{k}[j]}{\operatorname{min}} \ \mathbb{E} \left[\mathbb{G}_k^i[j] \Big\vert \mathcal{J}^i_k[j], \delta_k^i[j]\right].
	\end{align*}
	From \eqref{eq: hop cost} we observe that if $j < L$ then
	\begin{align*}
	\mathbb{E} \left[\mathbb{G}_k^i[j] \Big\vert \mathcal{J}^i_k[j], \delta_k^i[j]\right] & = \mathbb{E} \left[\sum_{l=j}^{j+1}  {-2 e_{k+1\vert k+1}^i[1]}^\top \Gamma_{k+1}^i \tilde{x}_{k+1\vert k+1}^i[l]+ {\tilde{x}_{k+1\vert k+1}^i[l]}^\top \Gamma_{k+1}^i \tilde{x}_{k+1\vert k+1}^i[l]  +  \lambda[j]\delta_k^i[j] \Bigg\vert \mathcal{J}^i_k[j], \delta_k^i[j] \right]\\
	& + \mathbb{E} \left[\sum_{l=j+2}^{L}  {-2 e_{k+1\vert k+1}^i[1]}^\top \Gamma_{k+1}^i \tilde{x}_{k+1\vert k+1}^i[l]+ {\tilde{x}_{k+1\vert k+1}^i[l]}^\top \Gamma_{k+1}^i \tilde{x}_{k+1\vert k+1}^i[l] \Bigg\vert \mathcal{J}^i_k[j]\right].
	\end{align*}
	We define the value of information as the gain in the cost when a measurement is successfully sent as opposed to
	when it is blocked, i.e., 
	\begin{align*}
	\numberthis\label{eq: value of information def}
	\textbf{dVoI}_{k}^i[j] &= \mathbb{E} \left[\mathbb{G}_k^i[j] + \mathbb{H}_{\pi, k+1}^i[j+1] \Big\vert \mathcal{J}^i_k[j], \delta_k^i[j]=1\right] - \mathbb{E} \left[\mathbb{G}_k^i[j] + \mathbb{H}_{\pi, k+1}^i[j+1]\Big\vert \mathcal{J}^i_k[j], \delta_k^i[j]=0\right].
	\end{align*}
	Before giving the general expression we look at the special case $j=L$. It follows that 
	\begin{align*}
	\mathbb{E} \left[\mathbb{G}_k^i[L] \Big\vert \mathcal{J}^i_k[L], \delta_k^i[L]\right] & = \mathbb{E} \left[ {\tilde{x}_{k+1\vert k+1}^i[L]}^\top \Gamma_{k+1}^i \tilde{x}_{k+1\vert k+1}^i[L]  +  \lambda[j]\delta_k^i[L] \Big\vert \mathcal{J}^i_k[L], \delta_k^i[L] \right],
	\end{align*}
	where we used expressions \eqref{eq: mismatch error espansion} and the fact that
	\begin{align}
	\label{eq: zero cross product}
	\mathbb{E} \left[ {e_{k+1\vert k+1}^i[1]}^\top \Gamma_{k+1}^i \tilde{x}_{k+1\vert k+1}^i[j]\Big\vert \mathcal{J}^i_k[j], \delta_k^i[j] = 1 \right] - 	\mathbb{E} \left[ {e_{k+1\vert k+1}^i[1]}^\top \Gamma_{k+1}^i \tilde{x}_{k+1\vert k+1}^i[j]\Big\vert \mathcal{J}^i_k[j], \delta_k^i[j] = 0 \right]= 0,
	\end{align}
	and $\tilde{x}_{k \vert k}^i[j+1]$ is $\mathcal{J}^i_k[j]$ measurable. Recalling that $\mathbb{H}_{\pi, t}^i[L+1]=0, \forall t$, then
	\begin{align*}
	\textbf{dVoI}_{k}^i[L]& = -\left(\tilde{x}_{k \vert k}^i[L]\right)^\top A_i^\top \Gamma^i_{ k+1} A_i\tilde{x}_{k \vert k}^i[L] + \lambda[L]
	\end{align*}
	The corresponding optimal scheduling policy is
	\begin{align*}
	\begin{aligned}
	{\delta_k^i[L]}^\star = \begin{cases}
	1, \quad \text{if } \textbf{dVoI}_{k}^i[L]<0, \\
	0, \quad \text{otherwise}.
	\end{cases}
	\end{aligned} \Leftrightarrow \
	\begin{aligned}
	{\delta_k^i[L]}^\star = \begin{cases}
	1, \quad \text{if } \left(\tilde{x}_{k \vert k}^i[L]\right)^\top A_i^\top \Gamma_{ k+1} A_i\tilde{x}_{k \vert k}^i[L] > \lambda[L], \\
	0, \quad \text{otherwise}.
	\end{cases}
	\end{aligned}
	\end{align*}
	For an arbitrary scheduler $j<L$ the expression of the value of information is given by
	\begin{align*}
	\textbf{dVoI}_{k}^i[j]& =\mathbb{E} \left[\sum_{l=j}^{\min\left(j+1, L\right)} {-2 e_{k+1\vert k+1}^i[1]}^\top \Gamma_{k+1}^i \tilde{x}_{k+1\vert k+1}^i[l]+ {\tilde{x}_{k+1\vert k+1}^i[l]}^\top \Gamma_{k+1}^i \tilde{x}_{k+1\vert k+1}^i[l]  +  \lambda[j]\delta_k^i[j]\Bigg\vert \mathcal{J}^i_k[j], \delta_k^i[j]=1\right]\\
	&-\mathbb{E} \left[\sum_{l=j}^{\min\left(j+1, L\right)}{-2 e_{k+1\vert k+1}^i[1]}^\top \Gamma_{k+1}^i \tilde{x}_{k+1\vert k+1}^i[l]+ {\tilde{x}_{k+1\vert k+1}^i[l]}^\top \Gamma_{k+1}^i \tilde{x}_{k+1\vert k+1}^i[l]\Bigg\vert \mathcal{J}^i_k[j], \delta_k^i[j]=0\right]\\
	& + \mathbb{E} \left[ \mathbb{H}_{\pi, k+1}^i[j+1] \Big\vert \mathcal{J}^i_k[j], \delta_k^i[j]=1\right] - \mathbb{E} \left[\mathbb{H}_{\pi, k+1}^i[j+1]\Big\vert \mathcal{J}^i_k[j], \delta_k^i[j]=0\right].
	\end{align*}
	We look at the difference
	\begin{align*}
	&\mathbb{E} \left[\sum_{l=j}^{\min\left(j+1, L\right)} {-2 e_{k+1\vert k+1}^i[1]}^\top \Gamma_{k+1}^i \tilde{x}_{k+1\vert k+1}^i[l] \Bigg\vert \mathcal{J}^i_k[j], \delta_k^i[j]=1\right]-\mathbb{E} \left[\sum_{l=j}^{\min\left(j+1, L\right)}{-2 e_{k+1\vert k+1}^i[1]}^\top \Gamma_{k+1}^i \tilde{x}_{k+1\vert k+1}^i[l]\Bigg\vert \mathcal{J}^i_k[j], \delta_k^i[j]=0\right]\\
	& \overset{\eqref{eq: zero cross product}}{=} \mathbb{E} \left[{-2 e_{k+1\vert k+1}^i[1]}^\top \Gamma_{k+1}^i \tilde{x}_{k+1\vert k+1}^i[j+1] \Bigg\vert \mathcal{J}^i_k[j], \delta_k^i[j]=1\right]-\mathbb{E} \left[{-2 e_{k+1\vert k+1}^i[1]}^\top \Gamma_{k+1}^i \tilde{x}_{k+1\vert k+1}^i[j+1]\Bigg\vert \mathcal{J}^i_k[j], \delta_k^i[j]=0\right]\\
	& \overset{\eqref{eq: mismatch error espansion j>1}}{=} \mathbb{E} \left[{-2 e_{k+1\vert k+1}^i[1]}^\top \Gamma_{k+1}^i \left(A_i \tilde{x}_{k \vert k}^i[j] + \left(1-\delta_k^i[j+1]\right)A_i\tilde{x}_{k \vert k}^i[j+1]\right) \Bigg\vert \mathcal{J}^i_k[j], \delta_k^i[j]=1\right] \\
	& \quad - \mathbb{E} \left[{-2 e_{k+1\vert k+1}^i[1]}^\top \Gamma_{k+1}^i \left(\left(1-\delta_k^i[j+1]\right)A_i\tilde{x}_{k \vert k}^i[j+1]\right) \Bigg\vert \mathcal{J}^i_k[j], \delta_k^i[j]=0\right] \\
	& = \mathbb{E} \left[{-2 e_{k+1\vert k+1}^i[1]}^\top \Gamma_{k+1}^i \left(A_i \tilde{x}_{k \vert k}^i[j] + \left(1-\delta_k^i[j+1]\right)A_i\tilde{x}_{k \vert k}^i[j+1]\right) \Bigg\vert \mathcal{J}^i_k[j]\right] \\
	& \quad - \mathbb{E} \left[{-2 e_{k+1\vert k+1}^i[1]}^\top \Gamma_{k+1}^i \left(\left(1-\delta_k^i[j+1]\right)A_i\tilde{x}_{k \vert k}^i[j+1]\right) \Bigg\vert \mathcal{J}^i_k[j]\right]=0.
	\end{align*}
	Where we used the fact that $\tilde{x}_{k \vert k}^i[j]$ is $\mathcal{J}^i_k[j]$ measurable. Using the previous equality it then follows that
	the value of information is given by
	\begin{align*}
	\textbf{dVoI}_{k}^i[j]& =\mathbb{E} \left[\sum_{l=j}^{\min\left(j+1, L\right)} {\tilde{x}_{k+1\vert k+1}^i[l]}^\top \Gamma_{k+1}^i \tilde{x}_{k+1\vert k+1}^i[l]  +  \lambda[j]\delta_k^i[j]\Bigg\vert \mathcal{J}^i_k[j], \delta_k^i[j]=1\right]\\
	&\ -\mathbb{E} \left[\sum_{l=j}^{\min\left(j+1, L\right)} {\tilde{x}_{k+1\vert k+1}^i[l]}^\top \Gamma_{k+1}^i \tilde{x}_{k+1\vert k+1}^i[l]\Bigg\vert \mathcal{J}^i_k[j], \delta_k^i[j]=0\right]\\
	& \ + \mathbb{E} \left[ \mathbb{H}_{\pi, k+1}^i[j+1] \Big\vert \mathcal{J}^i_k[j], \delta_k^i[j]=1\right] - \mathbb{E} \left[\mathbb{H}_{\pi, k+1}^i[j+1]\Big\vert \mathcal{J}^i_k[j], \delta_k^i[j]=0\right] \\
	&\overset{\eqref{eq: mismatch error espansion}}{=} \lambda[j] + \mathbb{E} \left[ \mathbb{H}_{\pi, k+1}^i[j+1] \Big\vert \mathcal{J}^i_k[j], \delta_k^i[j]=1\right] - \mathbb{E} \left[\mathbb{H}_{\pi, k+1}^i[j+1]\Big\vert \mathcal{J}^i_k[j], \delta_k^i[j]=0\right].
	\end{align*}
	\subsubsection*{Part 2}
	The optimal triggering policy provided above depends on the
	variable $\rho_k^i[j] \triangleq \mathbb{E} \left[ \mathbb{H}_{\pi, k+1}^i[j+1] \Big\vert \mathcal{J}^i_k[j], \delta_k^i[j]=1\right] - \mathbb{E} \left[\mathbb{H}_{\pi, k+1}^i[j+1]\Big\vert \mathcal{J}^i_k[j], \delta_k^i[j]=0\right]$. Although $\rho_k^i[j]$ can be computed with an arbitrary accuracy by solving recursively the optimality equation in \eqref{eq: value function of scheduler}, its computation is expensive \cite{soleymani2018value}. For its approximation we use the Rollout algorithm \cite{bertsekas2012dynamic} with baseline policy $\bar{\pi}$ given by ${\delta}_{k+1:T-1}[l]=1$, $l>j$.
	For $j\neq L$ we must find an approximation for the value function in order to compute the value of information
	\begin{align*}
	\textbf{dVoI}_{k}^i[j]& = \lambda[j]+ \mathbb{E} \left[ \mathbb{H}_{\pi, k+1}^i[j+1] \Big\vert \mathcal{J}^i_k[j], \delta_k^i[j]=1\right] - \mathbb{E} \left[\mathbb{H}_{\pi, k+1}^i[j+1]\Big\vert \mathcal{J}^i_k[j], \delta_k^i[j]=0\right].
	\end{align*}
	The case of $j=L-1$ reads as
	\begin{align*}
	\textbf{dVoI}_{k}^i[L-1]& = \lambda[L-1]+ \mathbb{E} \left[ \mathbb{H}_{\pi, k+1}^i[L] \Big\vert \mathcal{J}^i_k[L-1], \delta_k^i[L-1]=1\right] - \mathbb{E} \left[\mathbb{H}_{\pi, k+1}^i[L]\Big\vert \mathcal{J}^i_k[L-1], \delta_k^i[L-1]=0\right]\\
	& = \lambda[L-1]+ \mathbb{E} \left[ \mathbb{G}_{k+1}^i[L] \Big\vert \mathcal{J}^i_k[L-1], \delta_k^i[L-1]=1\right] - \mathbb{E} \left[\mathbb{G}_{k+1}^i[L]\Big\vert \mathcal{J}^i_k[L-1], \delta_k^i[L-1]=0\right]
	\end{align*}
	Using the same arguments as before and with the assumption that $\delta_{k:T-1}[l]=1$ and $\delta_{k+1:T-1}[j]=1$ it can be concluded that
	\begin{align*}
	\textbf{dVoI}_{k}^i[L-1]& =-\left(\tilde{x}_{k \vert k}^i[L-1]\right)^\top \left({A_i^2}\right)^\top \Gamma^i_{ k+2} {A_i^2}\tilde{x}_{k \vert k}^i[L-1] + \lambda[L-1].
	\end{align*}
	Iterating the procedure we find the general expression 
	\begin{align*}
	\textbf{dVoI}_{k}^i[j]& =-\left(\tilde{x}_{k \vert k}^i[j]\right)^\top \left({A_i^{L+1-j}}\right)^\top \Gamma^i_{ k+(L+1-j)} A_i^{L+1-j}\tilde{x}_{k \vert k}^i[j] + \lambda[j].
	\end{align*}

\end{document}